\newcommand{\ifacm}[1]{}
\newcommand{\ifusenix}[1]{#1}

\ifacm{
\documentclass[sigconf]{acmart}

\renewcommand\footnotetextcopyrightpermission[1]{} 
\setcopyright{none}

\settopmatter{printacmref=false, printccs=false, printfolios=true}

\acmDOI{}

\acmISBN{}

\acmConference[Submitted for review to SIGCOMM]{}
\acmYear{2026}
\copyrightyear{2026}

\acmPrice{}

}

\ifusenix{

\documentclass[letterpaper,twocolumn,10pt]{article}
\usepackage{usenix2019_v3}
  
}

\usepackage{tikz}
\usepackage{amsmath}
\usepackage{subfig}
\usepackage{wrapfig}
\usepackage{filecontents}
\usepackage{caption}
\usepackage{xspace}
\usepackage{xcolor}
\usepackage{multirow}
\usepackage{graphicx}
\usepackage{algorithm}
\usepackage{algpseudocode}
\usepackage{amsthm}
\usepackage{outline}
\usepackage{yhmath}

\newcommand{\cut}[1]{}
\newcommand{\ie}{\emph{i.e.}\xspace}
\newcommand{\etc}{\emph{etc.}\xspace}
\newcommand{\eg}{\emph{e.g.}\xspace}
\newcommand{\vs}{\emph{vs}\xspace}

\newcommand{\Fig}[1]{Figure \ref{fig:#1}\xspace}

\newcommand{\ngs}[1]{\textcolor{red}{#1}}

\newcommand{\nop}[1]{}
\newcommand{\Para}[1]{\vspace{4pt}\noindent\textbf{\textit{#1}}}

\newcommand{\ctp}{causally-triggered request\xspace}
\newcommand{\ctps}{causally-triggered requests\xspace}
\newcommand{\Ctps}{Causally-triggered requests\xspace}

\newcommand{\cp}{causal pair\xspace}
\newcommand{\cps}{causal pairs\xspace}
\newcommand{\Cps}{Causal pairs\xspace}
\newcommand{\CPs}{Causal Pairs\xspace}

\newcommand{\IPG}{IPG\xspace}
\newcommand{\IPGs}{IPGs\xspace}
\newcommand{\IBG}{IBG\xspace}

\def\compactify{\itemsep=0pt \topsep=0pt \partopsep=0pt \parsep=0pt \leftmargin=12pt}
\let\latexusecounter=\usecounter

\newcommand{\ct}{\small \tt}
\newcommand{\Sec}[1]{\S\ref{sec:#1}\xspace}
\newcommand{\App}[1]{Appendix \ref{app:#1}\xspace}
\newcommand{\Alg}[1]{Algorithm \ref{alg:#1}\xspace}

\newcommand{\forcameraready}[1]{}

\usepackage{listings}


\newcommand{\OurSystem}{{\sc Pirate}\xspace}
\newcommand{\oursystem}{{\sc Pirate}\xspace}

\begin{document}

\title{Passively Measuring Application Response
  Latency}


\author{Bhavana Vannarth Shobhana\textsuperscript{1}, Yen-lin Chien\textsuperscript{1}, Mark Doughten\textsuperscript{1}, Jonathan \\
Diamant\textsuperscript{2}, Badri Nath\textsuperscript{1}, Shir Landau Feibish\textsuperscript{2}, Srinivas Narayana\textsuperscript{1} \\
{\small \textsuperscript{1} Rutgers university \quad \textsuperscript{2} University of Haifa}}

\ifusenix{  
  \maketitle
}



\begin{abstract}
%

Latency is a key indicator of Internet service
performance. Continuously and passively tracking
the performance of actual user requests enables
the operators of Internet services to quickly
adapt resource allocation, and the operators of
wide-area ISP networks to build a deeper
understanding of customer experience. Ideally,
such measurements should be possible from vantage
points intermediary to the client and server, and
avoid instrumenting either.

While the measurement of transport-layer
round-trip time (RTT) is well studied, we find
that transport RTT is a poor proxy for application
response latency in Web and RPC traffic widely
prevalent in modern service
architectures. Unfortunately, tracking response
latency directly at intermediate vantage points is
thwarted by pervasive encryption of application
headers, and increasingly, transport headers.

This paper presents a passive estimator for
application response latency. \OurSystem leverages
the time gap between {\em causal packet
pairs}---two requests such that the response to
the first triggered the second. We introduce
techniques to identify causal packet pairs even in
the presence of multiple pipelined in-flight
requests. In experiments on real browsers
targeting top websites from a popular website
ranking, \OurSystem delivers more accurate
estimates of the client-visible response latency
than the most accurate RTT estimators. Since
\OurSystem only relies on packet timing and only in
one direction, it can operate over encrypted
traffic and under asymmetric routing.

\end{abstract}

\ifacm{
  \maketitle
}

\section{Introduction}
\label{sec:introduction}

Latency is a key indicator of the performance and
quality of interactive Internet services.
It is well known that smaller client-visible
latencies drive better user
engagement~\cite{gentleaggression-sigcomm13,
  cost-of-latency-2009, marissa-mayer-web20,
  controlled-experiments-kdd07,
  deloitte-latency-study,
  google-mobile-site-speed-playbook}.
%
%
Accurate measurements of client-visible
application latency can enable better content
placement~\cite{whyhigh-imc09,
  latency-estimation-networks08}, higher-quality
load balancing and
provisioning~\cite{prequal-nsdi24, cheetah-nsdi20,
  c3-nsdi15, knapsack-conext25}, and deeper user
experience monitoring in wide-area ISP
networks~\cite{persistent-interdomain-congestion-sigcomm18,
  sla-monitoring-sigcomm07}.
Continuous measurement is crucial, since
performance on the same connection may change over
time due to variable network
conditions~\cite{latency-variation-conext16,
  millisampler-imc22, incast-measurement-imc24,
  pingmesh-sigcomm15} and server
performance~\cite{seer-asplos19, heracles-isca15}.
Ideally, ``always-on'' monitoring should measure
user traffic directly and {\em passively} without
client or server
instrumentation~\cite{everflow-sigcomm15,
  snap-nsdi11}. Unlike approaches that actively
issue measurement probes, passive approaches do
not consume extra network bandwidth or trigger
alerts that signal volumetric attacks, even when
continuously measuring many endpoints at
scale~\cite{iplane-osdi06,
  latency-estimation-networks08}.

\Para{Application-layer Response latency.} In this
paper, we are interested in passively and
continuously measuring the time between when an
application-layer request is sent and when the
{\em last} byte of the corresponding response is
delivered to the client application, \ie the
per-object time-to-last-byte. We call this
quantity the application response latency or
simply {\em response latency.}
For RPC requests, response latency corresponds
to the completion time at the application
layer~\cite{tales-of-the-tail-socc14,
  fast-rpc-nsdi19}.  For web pages, the response
latencies of objects on the page are crucial
components in determining quality-of-experience
metrics such as the largest contentful
paint~\cite{qoe-of-web-users-ccr16,
  core-web-vitals, largest-contentful-paint,
  cloudflare-site-speed-matters}.

At first glance, response latency appears closely
related to the transport-layer Round-Trip Time
(RTT), whose passive measurement is widely studied
in the networking community
(\Sec{motivation-background}). However, response
latency is conceptually distinct from
transport-layer RTT. A server may return a
transport-layer acknowledgment well before it
returns the first byte of a response. Responses
may span multiple packets; however, the
transport-layer RTT accounts at most for the first
one. The time to the first byte of a response is
often a bad proxy for application-layer
performance, since it does not capture what must
still happen after the first byte is received. RPC
clients cannot typically progress without a full
response object. On the web, servers may
%
%
provide partial, ``early hint''
responses~\cite{http-early-hints} to ask a client
to load supporting resources, while still putting
together and transmitting the actual
response~\cite{measuring-ttfb-cloudflare-blogpost}.

Our goal is to design techniques that passively
and continuously measure response latency for
ongoing connections at an intermediate measurement
vantage point which can be unilaterally deployed
without instrumenting either clients or servers.
This would be straightforward to do if we could
match application-layer requests with their
responses, \eg by inspecting the HTTP headers.
However, encryption at the application layer is
pervasive on the wide-area web and also inside
compute clusters~\cite{ssl-statistics,
  cilium-mTLS, istio-mTLS, linkerd-mTLS}. Modern
network-layer security goes even further. With the
increasing adoption of encrypted network-layer
tunnels~\cite{wireguard, ipsec-esp,
  vpn-statistics}, the transport headers too
are hidden. Hence, any design seeking wide
practical applicability cannot rely on application
or transport headers.

Additionally, the compute and memory overheads of
measurement should be small enough to make it
tractable for a content provider or a wide-area
ISP to deploy it on a software middlebox. For
example, an ISP may piggyback the algorithm on a
middlebox that already processes packets for other
reasons. Approaches that store per-packet data
(\eg traces) or attempt to decrypt packets will be
found intractable.

\Para{A response latency estimator
  (\Sec{design}).} We introduce a continuous
passive estimator of response latency which is
agnostic of application and transport headers. We
build on three key ideas.

Our first idea is to leverage the closed-loop
  nature of Internet protocols and applications.
%
%
When application requests depend on the contents
of prior responses (for example, when a web object
embeds other objects), the reception of a response
triggers the generation of the subsequent
request. Flow-controlled applications mandate that
new requests are only transmitted when prior
responses arrive. Such closed-loop request
transmission behavior enables estimating the
response latency {\em by proxy:} a measurement
vantage point can approximate the response latency
by the time delay between a request and a
subsequent request triggered by the former's
response. We call the latter request a {\em \ctp,}
and the pair of requests a {\em \cp.} 
This method does not require matching
application-layer requests with responses.
%
%
%
%
%
Latency-sensitive services are known
to exhibit cross-request
dependencies~\cite{polaris-nsdi16,dapper-google-tracing10,
  alibaba-microservices-socc21, seer-asplos19,
  grpc-flow-control} and flow
control~\cite{memcache-facebook-nsdi13,http2-multiplexing-flow-control,
  quic-multiplexing-flow-control}, making causal
pairs widely prevalent in practice.


It is not obvious how to identify causal pairs. At any
given time, many concurrent requests may be in flight.
Connection persistence and stream multiplexing is
standardized and widely deployed in the HTTP
protocols~\cite{http-2-multiplexing,
  http-3-multiplexing, http-1.1-persistent}. 
Two consecutive requests arriving at a vantage point
from the same connection do not necessarily form a
causal pair.

Our second key idea is to leverage the time gaps
between packet arrivals to identify causal
pairs. Once a client has sent a few requests, its
transmission of subsequent requests is typically
blocked by dependencies or flow control. Hence,
causally-triggered requests arrive after a time
gap that is longer than the inter-arrival times of
the packets just prior. We call this the {\em
  prominent packet gap.} By choosing an
appropriate threshold on packet inter-arrival
times, a vantage point can classify any packet
arriving with a time gap exceeding this threshold
as a \ctp. The time interval between two
consecutive \ctps provides one estimate of the
response latency.

But what should the time threshold be set to?  Is
it possible to make this setting robust across
deployments (\eg wide area \vs data center),
network load, transport dynamics (\eg TCP
flavors), and applications (\eg web \vs key-value
store)?

Our third key idea is as follows: While a single
prominent packet gap is subject to (unknown)
network, application, and server conditions,
observing packet inter-arrivals {\em over a period
  of time} provides a more robust picture. In this
paper, we show it is possible to leverage the
probability distribution of inter-arrival times to
identify prominent packet gaps, specifically by
computing the modes of the distribution. We devise
a lightweight method that uses a per-connection
histogram to estimate the average response latency
for the (configurable) time period over which the
distribution was measured.


We call our algorithm, a synthesis of the three 
ideas above, \OurSystem.
Our design does not require visibility into
application or transport headers.
\OurSystem maintains a constant, configurable
amount of memory per monitored connection, and
requires bounded computation per packet.

Moreover, since \OurSystem only relies on packet
inter-arrival times in one direction, it can
function under asymmetric routing, whose
prevalence is well established in the wide-area
Internet~\cite{paxson-thesis,
  reverse-traceroute-nsdi10,
  outbound-stub-traffic-engineering-ccr04} and in
data
centers~\cite{load-balancing-in-dcs-survey18}.
As a practical use case, \OurSystem can enable
real-time, performance-aware decisions in
high-speed layer-4 load balancers. Such load
balancers use direct server
return~\cite{ananta-sigcomm13, maglev-nsdi16} and
skip processing the responses, thus exhibiting
routing asymmetry.
\OurSystem can improve the performance of Internet
services by enhancing the layer-4 load balancer
unilaterally, without modifying clients, servers,
application software, or the network.


\Para{Results (\Sec{evaluation}).} We evaluate
\OurSystem on RPC and Web traffic under a mix of
network topologies, network characteristics, and
background traffic. In each case, we compare
\OurSystem against the client-measured application
response latency as well as the client-measured
transport-layer RTT.
Our Web traffic experiments use a headless Chrome
browser targeted at a uniform sample of 20 out of
the top-100 web sites from the Tranco site
ranking~\cite{tranco-list}. Our RPC experiments
use a client retrieving objects with sizes and
embeddings distributed according to the top 100
sites. The CPU allocated to the web server is
drawn from a trace measured in a real
cluster~\cite{borg-eurosys15}.

Within a single cluster, on RPC traffic, across
all monitored responses, \OurSystem achieves a
median relative error of 0.63\% relative to the
response latency, getting to within $\pm$15\% of
the response latency 90\% of the time. 
The transport-layer RTT underestimates response
latency by 30\% at the median.
On Web traffic, \OurSystem offers a median
relative error of 0.35\%, getting to
within 20\% of the response latency 80\% of the
time. The transport-layer RTT underestimates
response latency by 100\% at the median.
\OurSystem does not yet meet the ideal of
estimating response latency with 100\%
accuracies. However, it is more accurate than the
state-of-the-art method of relying on the
transport-layer RTT.
In our experiments, this high-level trend carries
over to wide-area networks, varying degrees of
background traffic, and time-varying cellular
networks.

We implemented a version of the \OurSystem
algorithm piggybacking on
Katran~\cite{katran-description}, an open-source
software layer-4 load balancer based on Maglev
hashing~\cite{maglev-nsdi16}.
A single 16-CPU-core system combining \OurSystem
and Katran in series retains the same
request-processing throughput as solo Katran at an
offered load of 80,000 requests per second and at
10,000 active connections. 
Latency-aware Katran cuts the 99th percentile
latency in our experimental setup by 37\% on
average across loads, relative to vanilla Katran.

\OurSystem has an important limitation. While
\OurSystem works well in the presence of multiple
{\em pipelined} in-flight requests, such as in
HTTP/1.1, it does not perform well for {\em
  multiplexed} protocols such as HTTP/2 and HTTP/3
QUIC.  Multiplexing allows responses to be sent
back in an order different from that of the
requests, hence a request appearing after a
prominent packet gap need no longer represent a
\ctp. Despite this limitation, measuring HTTP/1.1
remains practically important. CloudFlare, a
widely-deployed content delivery network, reports
that HTTP/1.1 API
traffic~\cite{cloudflare-http-api-usage} accounts
for a growing majority of web requests across
CloudFlare~\cite{cloudflare-api-traffic-explosion}.
Moreover, we do not know of any techniques that
measure the application response latencies under
pipelining, let alone multiplexing. This work
is a first step towards estimating response
latency for multiplexed protocols.

%
%

%

A shorter version of this paper appeared
previously at a workshop venue (citation elided
for anonymity). In comparison, this paper
incorporates new techniques to make estimation
more robust, significantly expands evaluations of
accuracy and overhead under realistic settings,
and designs a feedback controller that operates on
real-time measurements.
%

\nop{

\noindent (1) When requests were multiplexed
according to dependencies and flow control on the
same connection, \OurSystem achieved a median
error in response latency estimation of under 8\%
relative to the ground truth response latency
measured at the client. The error arises primarily
due to client-side processing, inflating the
request-to-response time relative to the
request-to-triggered-request time at the client
(which \OurSystem models with 0\% relative
error).

\noindent (2) When the client is artificially
restricted to keeping only a single request packet
in flight, \OurSystem's median error is 0\%. In
this regime, response latency equals TCP RTT, and
\OurSystem agrees closely with estimates obtained
from prior work on TCP RTT
estimation~\cite{tcptrace} and sample RTT
estimates from the Linux kernel network
stack~\cite{tcp-probe-tracepoint,
  ebpf-tracepoints-tutorial}.

Note that the ground-truth baselines above see
packets in both directions of traffic. \OurSystem
only observes client to server packets.

\noindent (3) The accuracy of \OurSystem is robust
to realistic degrees of packet loss and
reordering. \OurSystem provides high accuracy with
multiple applications including memcached.

\noindent (4) We integrated \OurSystem into
Katran~\cite{katran-description}, an open-source
Maglev load balancer. Latency-aware Katran cuts
the 99th percentile latency of the full web
workload in our experimental setup by $\sim$40\%
(across loads) relative to (unmodified)
Katran. Further, latency-aware Katran has much
less variability in the tail than Katran.

\noindent (5) On our experimental setup,
\OurSystem imposes an average delay of 346 ns in
the critical packet-processing path when
forwarding packets using XDP~\cite{xdp-conext18},
a fast packet-processing platform. The Katran load
balancer, also implemented in XDP) takes on
average 1114 ns to complete.

\noindent (5) Since \OurSystem relies on packet
timings (and not transport headers), it can
passively measure the response latencies of HTTP/3
connections over QUIC despite encrypted headers
and asymmetric routing.

In each case, we discuss the source of the errors
that \OurSystem makes. A previous shorter version
of this paper appeared at a workshop (citation
elided for anonymity). In comparison, this paper
adds several new techniques to make response
latency estimation more robust, including
algorithms for maintaining empirical distributions
and estimating average latencies from
them. Further, we have added new evaluations of
the accuracy and overhead of our techniques in
realistic environments. We have also demonstrated
a feedback controller (layer-4 load balancer)
relying on latency measurements.

}





\section{Background and Motivation}
\label{sec:motivation-background}

We first survey the literature on transport RTT
estimation, a problem closely related to measuring
application response latency. We also discuss why
it is nontrivial to generalize these techniques to
measuring application-layer response latency.

\label{sec:response-latency-and-rtt-measurement}


\Para{Passive estimation of transport-layer RTTs.}
%
%
The basic idea of passive transport RTT estimation
is matching data packets to their corresponding
transport-layer acknowledgments (ACKs), and
measuring the time difference between this pair of
packets.
The matching is made possible by connecting the
sequence numbers or timestamps on TCP data packets
to the corresponding information on the ACK
packets~\cite{tcptrace, dapper-sosr17,
  dart-sigcomm22, stretch-acks-ccr15,
  network-methods-passive-estimation-pam05,
  passive-tcp-connection-characteristics-infocom04}.
Implementing the basic approach requires
visibility into transport headers in the clear and
seeing packets in both directions.

An early approach from Jiang and
Dovrolis~\cite{passive-estimation-tcp-rtt-ccr02}
estimates RTT through the time interval between
the SYN and the ACK packet of the TCP 3-way
handshake, and also between small packet bursts
during the early rounds of slow start.
The syn-ack
estimator~\cite{latency-estimation-networks08,
  monitoring-encrypted-traffic-inm07} is usable
even under routing asymmetry. However, it cannot
measure the performance of connections beyond the
beginning. Connection performance can evolve with
time due to network and server performance
variability~\cite{latency-variation-conext16,
  millisampler-imc22, incast-measurement-imc24,
  pingmesh-sigcomm15, seer-asplos19,
  heracles-isca15}.
Some prior works estimate RTTs by computing
frequency spectra over the time series of packet
arrivals in one
direction~\cite{passive-one-way-networking10,
  network-methods-passive-estimation-pam05}.
However, computing frequency spectra requires
significant memory to hold the timings of all
packets over an interval. The computations also
tend to be complex to
tune in practice~\cite{stretch-acks-ccr15}.

Researchers have proposed the spin bit as a QUIC
protocol extension to passively measure
RTTs~\cite{three-bits-suffice-imc18,
  spin-bit-ietf,
  quic-spin-bit-deployment-imc23}. Since QUIC
encrypts both application and transport headers,
this effort comes closest in spirit to meeting our
measurement goals.  However, as we show next,
transport-layer RTT differs from the
application-layer response latency.

\begin{figure}
  \begin{center}
    \includegraphics[width=0.4\textwidth]{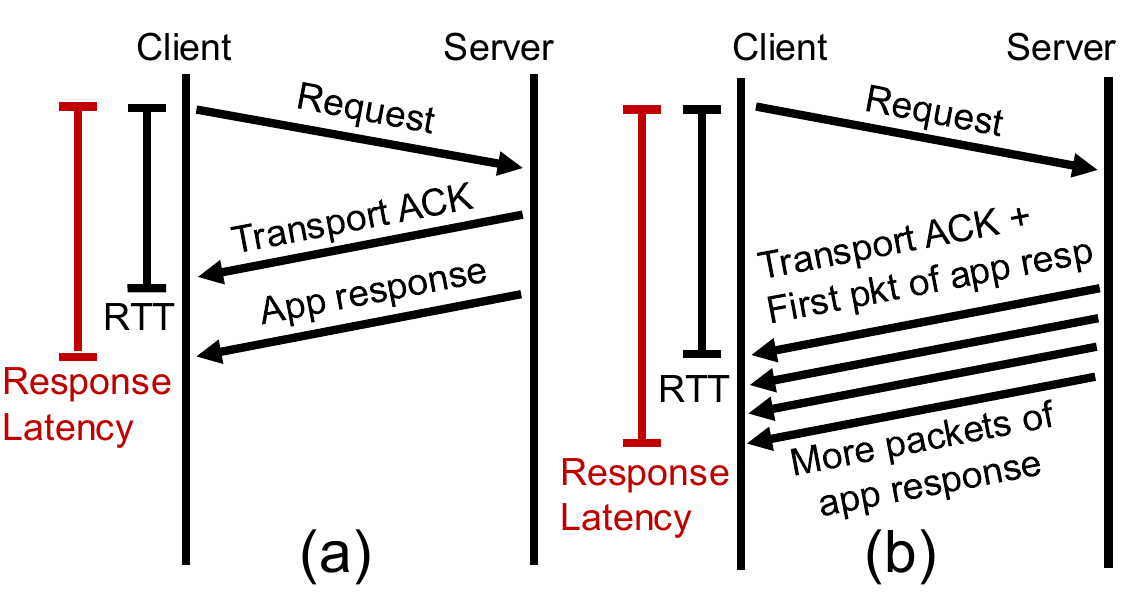}
    \caption{Transport-layer round-trip time (RTT)
      does not always represent the
      application-layer response latency.}
    \label{fig:response-latency-not-equal-tcp-rtt-schematic}
  \end{center}
\end{figure}

\Para{Response latency is dissimilar to the RTT.}
A high RTT implies a high response latency, but a low
RTT does not imply a low response latency.
The RTT and the response latency match exactly if
a server application sends exactly one response
packet that piggybacks the transport-layer ACK.
However, the server could transmit a
transport-layer ACK to a client prior to
transmitting any response data
(\Fig{response-latency-not-equal-tcp-rtt-schematic}(a)). Further,
receiving the end of a response larger than a
single packet will incur more time than the RTT
suggests
(\Fig{response-latency-not-equal-tcp-rtt-schematic}(b)).
When the client and server use a protocol with
pipelined requests (\eg HTTP/1.1), multiple
application requests may be transmitted in a
single request packet. The transport-layer ACK
corresponds at most to the first application-layer
response.
For all these reasons, the RTT is prone to
underestimating the response latency.

\begin{figure}
  \begin{center}
    \includegraphics[width=0.20\textwidth]{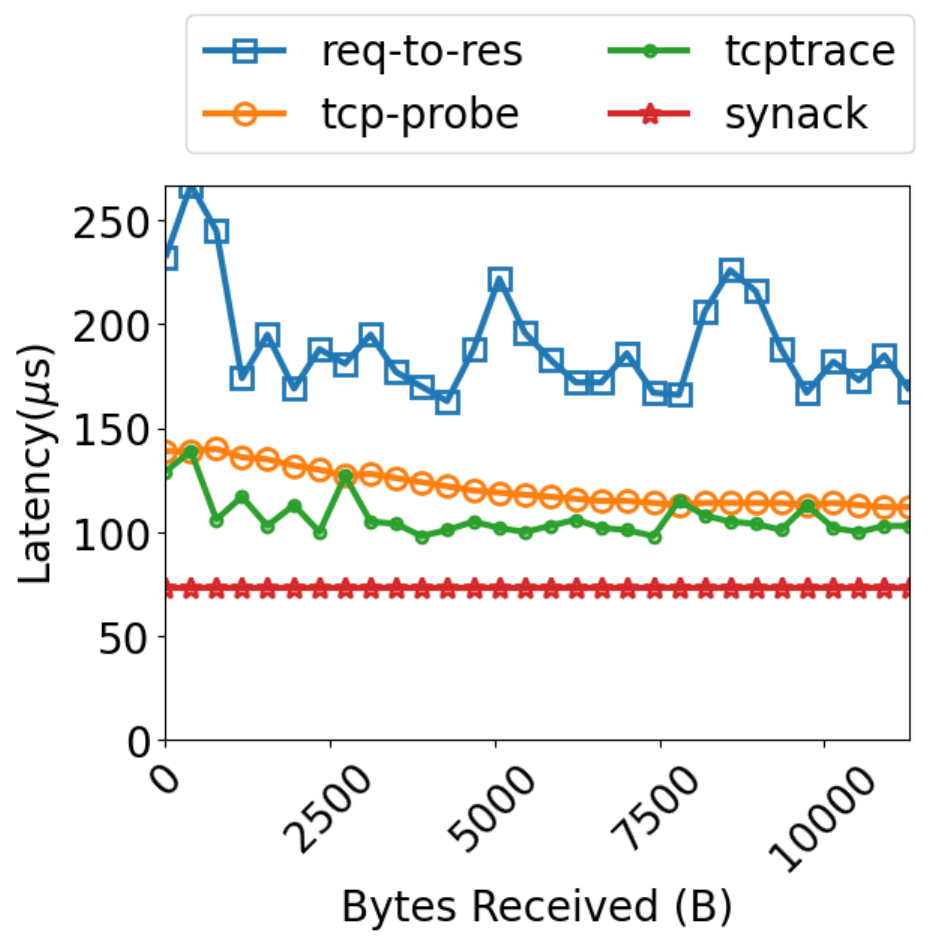}
    \includegraphics[width=0.25\textwidth]{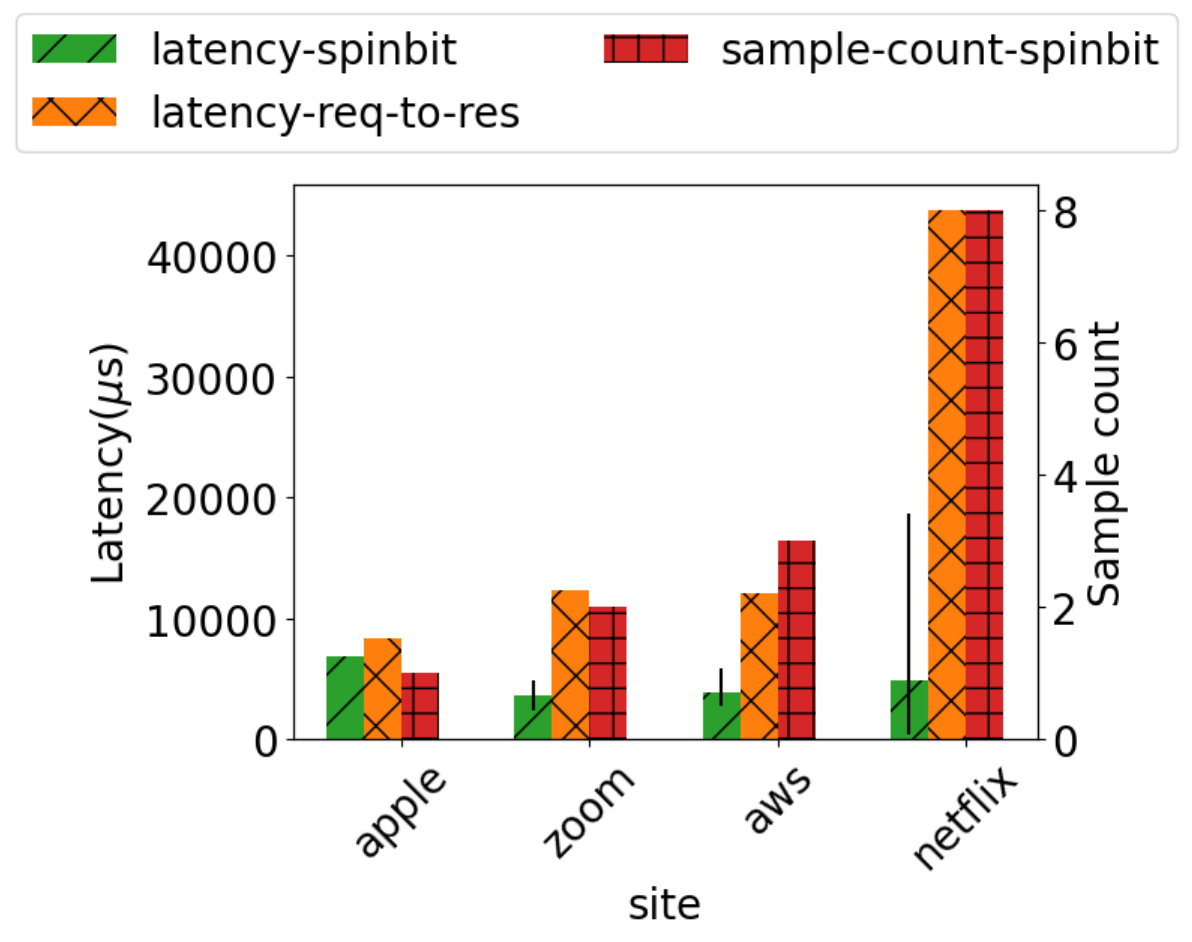}
    \caption{Transport round-trip times (RTTs) do
      not match response latency, with clients
      running (a) TCP HTTP/1.1 and (b) UDP
      QUIC. See discussion in
      \Sec{response-latency-and-rtt-measurement}.}
    \label{fig:response-latency-not-equal-tcp-rtt}
  \end{center}
\end{figure}

In our experiments, we find that RTT
underestimates response latency even when there is
only one request in flight with a small
application-layer response.  Our full
experimental setup is described in
\Sec{detailed-experimental-setup}.
In \Fig{response-latency-not-equal-tcp-rtt}(a), an
RPC client running over TCP maintains a single
outstanding request with the server. The figure
shows the time series of response latencies
(marked {\ct req-to-res}) against three techniques
that measure the RTT: (i) reading statistics from
the kernel network stack through {\ct
  tcp-probe}~\cite{tcp-probe-tracepoint}; (ii)
estimating RTT from a packet trace ({\ct
  tcptrace}~\cite{tcptrace}; and (iii) syn-ack
delay~\cite{passive-estimation-tcp-rtt-ccr02}.
Not only does the RTT underestimate the response
latency, but it is also uncorrelated with it (we
show more extensive experiments in
\Sec{evaluation}). Moreover, underestimation is
not a characteristic of the transport protocol
(TCP).  We conducted a similar experiment using 
the lsquic QUIC client library~\cite{litespeed-quic} on 
real web pages from a top-100 website ranking, 
operating over UDP/QUIC with the spin-bit protocol 
extension~\cite{three-bits-suffice-imc18}.
\Fig{response-latency-not-equal-tcp-rtt}(b) shows
that RTT and response latency diverge, especially
when there are multiple packets in a single
response. The error bars show the minimum and
maximum RTTs reported.

\Para{Measuring application performance directly
  from packet traces.}  Given the emergence of
encrypted transports on the Internet, some prior
work aims to passively infer quality of experience
metrics specifically for web
applications~\cite{qoe-from-encrypted-traffic-ifip20,
  time-series-modeling-web-qoe-encrypted-traffic,
  deployable-models-qoe-from-encrypted-traffic-tnsm21}.
These methods employ machine learning and
prediction over a sequence of connection features
accumulated in time to estimate an overall metric
of website experience, \eg~\cite{pagespeed}. In
the process, they compose the response latencies
of multiple objects in service of a single
aggregated performance metric per connection. In
contrast, our work focuses on continuously
measuring object response latencies over the
lifetime of a connection. This notion is general
across the Web and RPC traffic, and enables
real-time action (\eg rerouting) to benefit
ongoing connections. In summary, we are aware of
no techniques to passively measure
application-layer response latencies for encrypted
connections on an ongoing basis.

\nop{
There is also significant prior work on passively
measuring the latency of specific segments of a
network, \eg~\cite{lda-sigcomm09, rli-sigcomm10,
  millisampler-imc22, incast-measurement-imc24} and
algorithms to estimate latencies over end-to-end
network paths using previously measured
delays,~\eg~\cite{gnp-infocom02,
  structural-latency-imc06}. Plenty of active
approaches exist to measure end-to-end network latency
characteristics, \eg~\cite{sla-monitoring-sigcomm07,
  software-defined-latency-monitoring-pam15,
  tcp-congestion-signatures-imc17}. This paper seeks
continuous passive measurement techniques for
end-to-end response latency. 

In summary, we believe that passive continuous
measurement of response latencies under modern
transports and practical deployment constraints
is as yet unsolved.
}





\section{Designing a Response Latency Estimator}
\label{sec:design}

\newcommand{\efficientmodeupdatealgorithm}{
  \begin{algorithm}
  \begin{algorithmic}[1]
    %
    \Require{New observation of an \IPG $g$}
    %
    %
    \Require{A representation of the empirical
      probability distribution, $M$, with $N$
      modes. For each $1 \leq i \leq N$, the $i^{th}$
      mode is a tuple $(min, max, count, sum)$,
      denoting the minimum, maximum, count, and sum of
      all the \IPGs observed within the mode.}
    \Ensure{$M$ is updated with the additional \IPG
      $g$}
    \Function{UpdateModes}{\IPG $g$}
    \For{$m \in M.get\_modes():$}
    \Statex \Comment{Modes traversed in ascending order}
    \State $left = m.get\_min()$
    \State $right = m.get\_max()$
    \If{$left - \epsilon \leq g \leq right + \epsilon$}
    \Statex \Comment{IPG $g$ lies within or proximal to mode $m$}
    \State {\sc AddGapToMode}($g, m$)
    \If{$left - \epsilon \leq g \leq left$}
    \Statex \Comment{$g$ is proximal from below}
    \State $m.set\_min(g)$
    \State {\sc ConsiderMergeModes}($m.get\_prev(), m$)
    \ElsIf{$right \leq g \leq right + \epsilon$}
    \Statex \Comment{$g$ is proximal from above}
    \State $m.set\_max(g)$
    \State {\sc ConsiderMergeModes}($m, m.get\_next()$)
    \EndIf
    \State \Return
    \EndIf
    \EndFor
    \If{$M$ has fewer than $N$ initialized modes}
    \Statex \Comment{Insert singleton mode containing $g$}
    \State {\sc AddMode}$(M, g)$
    \Else
    \State discarded += 1
    \EndIf
    \EndFunction
  \end{algorithmic}      
  \caption{Maintaining a small number of modes $N$ from
    the empirical probability distribution of \IPGs.}
  \label{alg:efficient-mode-update}
  \end{algorithm}
}

\newcommand{\packetgapalgorithm}{
  \begin{algorithm}
\begin{algorithmic}[1]
  \Require{Fixed threshold on inter-packet gaps, $\delta$}
  \Require{Timestamp of the current packet's arrival, $now$}
  \Require{The last time a new batch arrived for flow $f$, $f.time\_last\_batch$}
  \Require{The last time a packet arrived for flow $f$, $f.time\_last\_pkt$}
  \Ensure{An estimate of flow $f$'s reponse latency, $\hat{T}_{LB}$, if a new sample is produced, else $undef$}
  \State $\hat{T}_{LB} \gets undef$ 
  \If{$now - f.time\_last\_pkt > \delta$}
    \Comment{New batch: record response latency}
    \State $\hat{T}_{LB} \gets now - f.time\_last\_batch$ 
    \State $f.time\_last\_batch \gets now$ 
  \EndIf
  \State $f.time\_last\_pkt \gets now$ 
  \State \Return $\hat{T}_{LB}$
\end{algorithmic}  
\caption{Track \ctps using a fixed time threshold
  $\delta$ at the vantage point.  The algorithm is
  executed upon receiving each packet of a flow $f$.}
\label{alg:fixed-timeout-detection}
\end{algorithm}
}

\newcommand{\packetgapalgorithmold}{
\begin{algorithm}
  \DontPrintSemicolon
  \KwIn{Fixed threshold on inter-packet gaps, $\delta$}
  \KwIn{Timestamp of the current packet's arrival, $now$}
  \KwIn{The last time a new batch arrived for flow $f$,
    $f.time\_last\_batch$}
  \KwIn{The last time a packet arrived for flow $f$,
    $f.time\_last\_pkt$}
  \KwOut{An estimate of flow $f$'s round trip time,
    $\hat{T}_{LB}$, if a new sample is produced, else $undef$}
  $\hat{T}_{LB} = undef$ \;
  \If{$now - f.time\_last\_pkt$ $>$ $\delta$}{
    \Comment{New batch: record response latency.}
    $\hat{T}_{LB} = now - f.time\_last\_batch$ \;
    $f.time\_last\_batch = now$ \;
  }
  $f.time\_last\_pkt = now$ \;
  \KwRet{$\hat{T}_{LB}$}
  \caption{Track causally-triggered
    transmissions through a fixed timeout to identify new
    batches of packets, executed at LB upon receiving each
    packet of flow $f$.}
  \label{alg:fixed-timeout-detection}
\end{algorithm}
}

This section introduces an algorithm to measure
response latency passively and continuously for
interactive applications. We assume that the
vantage point of measurement lies on the path from
the client to the server.

\subsection{\CPs}
\label{sec:causal-idea}

Instead of measuring the time interval between a
request and its response, our first key idea is to
measure a proxy time interval---the time between a
request and another request transmitted by the
client due to the reception of the former's
response. We call the latter packet a {\em \ctp},
and the pair of packets a {\em \cp}
(\Fig{causal-pairs-and-errors}).

\Ctps are prevalent in practice due to several
reasons: (1) Cross-request dependencies:
A client application may issue a subsequent
request only when responses to previous requests
have been received and processed by the
client. For example, web clients and RPC clients
issue follow-up requests to fetch objects based on
previous responses.
(2) Flow control: Clients frequently cap the
number of application or transport data that are
outstanding at the server. Web browsers, memcached
clients, and RPC client libraries are all known to
subject clients to such flow
control~\cite{http2-multiplexing-flow-control,
  quic-multiplexing-flow-control,
  grpc-performance-best-practices}.
(3) Acknowledgments:  A response may trigger an
acknowledgment at the transport or application
layer.

\begin{figure}
  \begin{center}
    \includegraphics[width=0.4\textwidth]{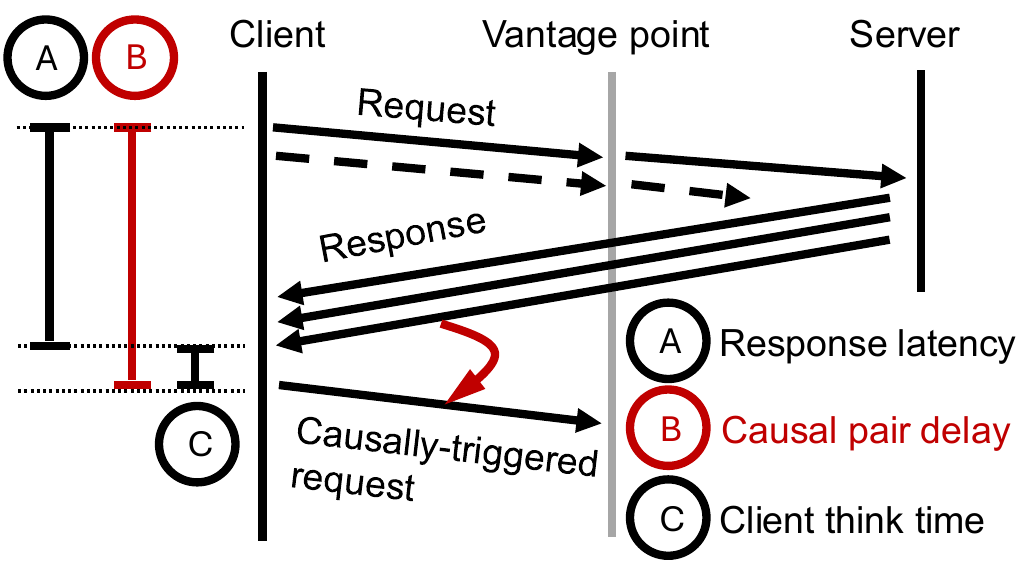}
    \caption{The time delay between a \cp
      (\Sec{causal-idea}) estimates the response
      latency of the first request in the
      pair. However, latency is overestimated by
      the time it takes the client to generate the
      triggered request (labeled the {\em client
        think time}). }
    \label{fig:causal-pairs-and-errors}
  \end{center}
\end{figure}

There are two caveats to using \cps to estimate
response latency.
First, a client application may incur additional
time to process a response prior to generating a
follow-up request, due to considerations such as
parsing or delays due to process or thread
scheduling.
We call this the {\em client think time}
(\Fig{causal-pairs-and-errors}).
%
Second, a response latency measurement is only
available for one request per \cp, not all the
requests transmitted between the two requests on
the same connection. 
%

If a client application maintains exactly one
request in flight, with the next request only
issued once the response to the first one arrives,
the two requests form a causal pair, and measuring
the time interval between the two requests
provides an estimate of the response latency. \Cps
are a generalization of syn-ack
estimation~\cite{passive-estimation-tcp-rtt-ccr02}. Like
syn-ack estimation, causal pair delays only
concern packets flowing in one direction, hence it
is possible to measure them under asymmetric
routing.

A key challenge arises when we go to settings
where multiple requests or packets are
concurrently in flight. Two consecutive requests
observed at a vantage point from the same
connection need not form a \cp. Accurate knowledge
of the client's ongoing window size (at the
transport or application layer) may help identify
packets that are not causally related with each
other, \eg if they belong to the same
window. However, inferring the window size
typically requires making specific assumptions on
the dynamics of the transport
protocol~\cite{passive-tcp-connection-characteristics-infocom04,
  dapper-sosr17}, a dependency that we seek to
avoid.



\subsection{Prominent Packet Gaps}
\label{sec:packet-gap-assumption}

Our second key idea is to leverage the timings of
packet arrivals to identify \cps when a client has
many concurrent requests in flight. 
The reasons that produce \ctps (cross-request
dependencies, flow control, acknowledgments)
result in clients transmitting a burst of
data in one shot, and then pausing request
transmission until a response arrives.
The sender side of transport often transmits in
bursts to leverage hardware batching optimizations
(\eg TSO~\cite{tso}), while receivers typically
optimize I/O and CPU by batching ACK
processing~\cite{tso, quic-ack-frequency-rfc}.
The prevalence of burst-then-pause behavior is
well documented: TCP senders that use window-based
transmission send bursts of packets, termed
flowlets~\cite{flare-flowlet-hotnets04,
  letflow-nsdi17}, separated by a pause.

We exploit the observation that the pause is
noticeably longer and more prominent than the
packet inter-arrival times in the burst, since the
former is subject to delays from the network and
server processing, while the latter is determined
only by the ability of the client to transmit
requests.
Hence, the first packet arriving after the pause
must be a \ctp.
The first packet of the previous burst is the
first packet of the \cp.
%
%
%


Given a fixed time threshold $\delta$, packets
which arrive with an inter-arrival time gap of
more than $\delta$ are considered to be a part of
different bursts. We could estimate the response
latency by the time between the first packets of
successive bursts  (shown in
\Alg{fixed-timeout-detection} in
\App{packet-gap-algorithm}).


How should we choose the time threshold $\delta$?
The threshold substantially impacts the accuracy of
our estimate. Choosing too large a $\delta$ will
miss legitimate causal pairs, only identifying
long idle periods in the connection. Choosing too
small a $\delta$ can make the estimation
vulnerable to noisy gaps between packets
transmitted within the ``burst,'' for example due
to scheduling or processing delays at the client's
application or transport layer.

Fundamentally, the duration of the pause depends on
several factors, such as the timing of writes from the
client application into the transport layer, the client
transport's scheduling of packet transmission (\ie flow
and congestion control), scheduling at the network
stack's traffic control layer or the
NIC~\cite{carousel-sigcomm17}, cross traffic competing
with the connection before packets arrive at the
vantage point, the network round-trip time, and server
processing delays. The combination of these factors
makes it unlikely that a fixed threshold $\delta$ can
work across different scenarios, or across time even
for a single connection in a specific
network. Approaches that select time thresholds for
flowlet-based network load
balancing~\cite{flare-flowlet-hotnets04, flare-ccr07,
  letflow-nsdi17, conga-sigcomm14} do so using the
differences between expected path latencies, or through
an empirically-tuned value that achieves a desired
balance among paths. These approaches are unsuitable
for a passive observer attempting to measure latency in
the first place.

The assumption of a prominent packet gap does not
hold if requests eligible for transmission are
paced rather than transmitted eagerly in a
burst. Pacing may happen either due to the
sender's transport, \eg~\cite{bbr-queue16,
  nimbus-sigcomm22} or a bottleneck link or
a policer~\cite{traffic-policing-google-sigcomm16,
  shaperprobe-imc11} preceding the vantage point. 
There are well known techniques to detect pacing
and policing using packet
timing~\cite{traffic-policing-google-sigcomm16,
  bbr-queue16}. Hence, one could detect and remove
paced connections from consideration of
measurement.
The landscape of sender transport behaviors on the
Internet is diverse~\cite{ccanalyzer-sigcomm24,
  inspector-gadget-ntma20, cc-census-measanal19},
leaving ample connections to measure.

\nop{We conjecture that latency-sensitive
  applications transmitting small amounts of
  bursty traffic may not be widely subject to
  pacing, capacity bottlenecks, or policing.}

\subsection{Choosing a Prominence Threshold}
\label{sec:estimating-packet-gap}

Our third key idea is that observing the timings
of packet arrivals {\em over time} can provide
meaningful clues to determining good values for
the pause time threshold $\delta$
(\Sec{packet-gap-assumption}). This observation is
inspired by prior work that motives a longitudinal
view of noisy data as a way to design robust
measurement and feedback
control~\cite{huygens-nsdi18,
  high-bdp-small-buffer-cc-ccr08,
  buffer-sizes-ccr05,
  control-theory-buffer-sizing-ccr05}.
Instead of producing an estimate instantaneously,
we can produce an average over a configurable time
epoch.

Specifically, given an empirical distribution of
inter-packet time gaps (\IPGs) observed over the
lifetime of a connection, the significant {\em modes}
of the probability distribution carry information about
specific events occurring on that connection. In
\Fig{ipg-modes}, we show an empirical distribution of
\IPGs in a simple experimental scenario where a TCP
client has a roughly constant response latency whenever
the response arrives, but experiences occasional packet
losses and idle periods (where no data is either
transmitted or received). The \IPG distribution
includes a batch of packets within a burst
(smallest-valued modes); \IPGs across bursts of
causally-dependent packets; loss timeouts (typically
set larger than response latencies); and idle periods
(the largest modes).

\begin{figure}
  \begin{center}
    \includegraphics[width=0.3\textwidth]{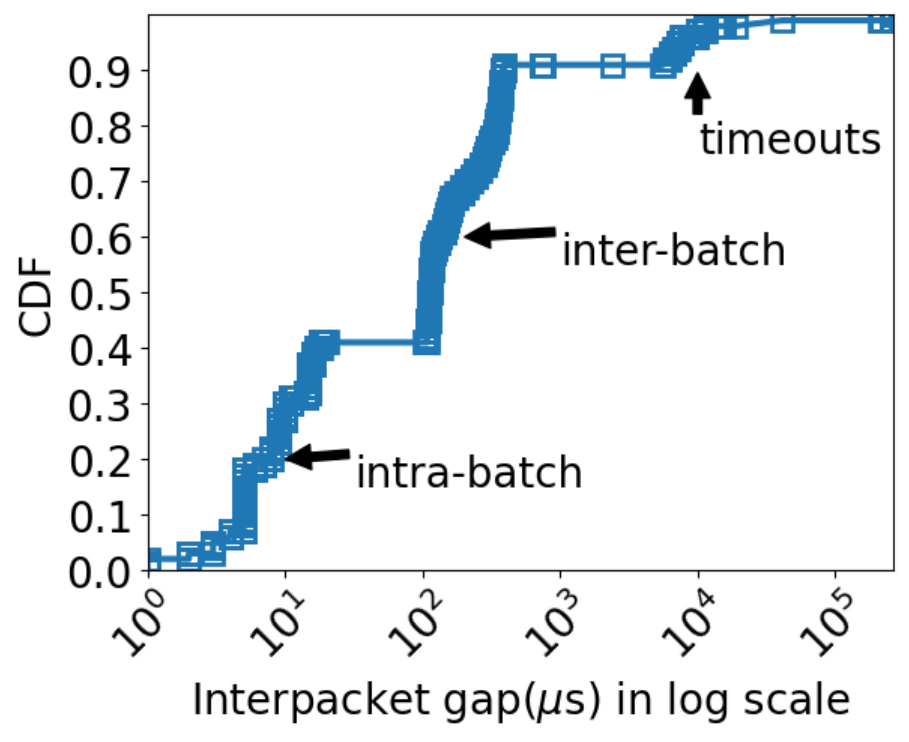}
    \caption{Modes in the empirical distribution
      of inter-packet gaps (IPGs). The modes carry
      useful information about phenomena of
      interest occurring over the measurement
      epoch (\Sec{estimating-packet-gap}).}
    \label{fig:ipg-modes}
  \end{center}
\end{figure}

\label{sec:empirical-probability-distribution}

It is possible to eliminate the modes
corresponding to retransmission timeouts and idle
periods through the knowledge of standard
parameters used in transport stacks. Timeouts are
publicly known for several standard TCP and QUIC
configurations~\cite{quic-probe-timeout,
  tcp-retransmission-timer-rfc,
  tcp-timeout-datacenters-sigcomm09}.
Hence, it is possible to eliminate the modes
corresponding to retransmission timeouts and idle
periods, leaving behind a {\em de-noised}
distribution of \IPGs, which only includes
inter-request gaps corresponding to either (i)
back-to-back transmissions without causal
dependencies, or (ii) \ctps with a prominent
packet gap.

\Para{Weighted-mode-sum estimator.}
We devise a simple estimation procedure which
assumes that (i) the empirical \IPG distribution
is replicated over each time interval that spans a
\cp during the epoch over which the distribution
is measured; and (ii) the largest mode in the
de-noised \IPG distribution is the \IPG preceding
the arrival of a \ctp. We call this \IPG the {\em
  inter-batch gap (\IBG)}. The average response
delay can be estimated by summing up the modes
smaller than the \IBG weighted by their frequency
relative to the \IBG. For modes $m_i$ (one of
which is the \IBG)

\vspace{0.1in}

$\widehat{response\_latency} = \displaystyle\sum_{m_i \leq IBG}
\frac{Pr(m_i)}{Pr(IBG)} * m_i$

\vspace{0.1in}

The intuition for the estimator is as follows. The
summation of all \IPGs over a \cp period gives us
an estimate of the response latency for the first
packet of the pair. According to assumption (i),
each \cp period consists of \IPGs whose values
equal the modes $m_i \leq IBG$. These modes occur
with frequencies proportional to their respective
probabilities $Pr(m_i)$. According to assumption
(ii), the \IBG mode occurs exactly once in a
single \cp period. Then, the ratio of the
probability of mode $m_i$ to that of the \IBG
tells us exactly how many occurrences of each
$m_i$ we should account for.

{\em An example.} Suppose the \IPG distribution
has three modes $m_1$ = 100 $\mu$s, $m_2$ = 150
$\mu$s, and $m_3$ = 250 $\mu$s (after de-noising),
with probabilities $Pr(m_1) = 0.4, Pr(m_2) = 0.2,
$ and $Pr(m_3) = 0.1$. We assume that $m_3$ (the
largest-valued mode) is the prominent packet gap
(\Sec{packet-gap-assumption}).  Corresponding to
each occurrence of $m_3$, there are $Pr(m_1) /
Pr(m_3) = 4$ occurrences of $m_1$ and $Pr(m_2) /
Pr(m_3) = 2$ occurrences of $m_2$. We estimate the
average response latency as $4*m_1 + 2*m_2 + m_3 =
950\ \mu$s.

Our use of \IPG distributions is distinct from prior
techniques that probe the network using packet trains
and use \IPG distributions to estimate bottleneck link
bandwidth, \eg~\cite{packet-dispersion-infocom01,
  bandwidth-estimation-network03}. The average response
latency is the sum of some number of packet gaps;
identifying which packet gaps to combine and in what
proportion is the core challenge of latency
measurement, which prior work does not tackle.

While \IPG distributions provide useful
information, computing distributions by
maintaining the full list of \IPGs for each active
connection (say, on a software middlebox) is
prohibitively memory-expensive. Naive methods to
maintain histograms can quickly get expensive,
too. With a sufficiently fine-grained time
resolution per bucket, say, 10 $\mu$s (the RTT
value in some data
centers~\cite{azure-accelnet-nsdi18,
  azure-latency-statistics}), an epoch length of
100 ms and a 2 byte counter per bucket of the
histogram, one connection would require (100 ms /
10 $\mu$s) * 2 bytes = 20 KBytes of memory. For a
vantage point that sees 10K active connections,
the memory consumption of the histogram touches
200 MBytes. This value is larger than the L2
caches on many server architectures, and typically
implies a slowdown in software packet-processing
performance. Using coarser time resolutions per
bucket sacrifices accuracy when response latencies
are small: the algorithm cannot distinguish modes
finer than the time resolution of each bucket.




\subsection{Maintaining Efficient Histograms}
\label{sec:efficient-ipg-distributions}

To maintain an \IPG distribution with memory
efficiency, we design an algorithm that maintains
a small number of buckets by dynamically varying
the resolution of each bucket according to the
\IPGs observed on that connection. Dynamic bucket
resolution has the advantages that (1) the
histograms of different connections may span
different ranges of \IPGs; (2) we can avoid the
overhead of counter bits for \IPGs that do not
occur; and instead, (3) we can focus the available
memory on maintaining more counter bits for \IPGs
actually observed over each connection.

Our algorithm works as follows. The histogram,
$M$, includes a configurable maximum number of
buckets (modes) $N$. Initially, all buckets are
uninitialized. Each bucket (after initialization)
specifies the minimum, maximum, count, and sum of
all the \IPGs observed within that bucket.
For each newly observed \IPG $g$, the algorithm
either (i) counts $g$ into an existing bucket
whose (min, max) interval covers $g$, or (ii) if
$g$ is ``close enough'' to an existing bucket,
extends the bucket or merges two buckets to
produce a new, larger bucket that now covers $g$,
or (iii) puts $g$ into its own new bucket if there
is an available uninitialized bucket, or (iv)
discards $g$.
%
%
Across our experiments, maintaining as few as
$N=10$ buckets per connection proves sufficient to
capture all \IPG distributions.
At the end of each epoch, the weighted mode
sum (\Sec{empirical-probability-distribution}) is
computed over the average mode values of $M$ to
emit a response latency averaged over the epoch.
Our complete algorithm is shown in
\Alg{efficient-mode-update} in \App{histogram-algo}.

The literature on streaming algorithms includes
other storage-efficient methods to maintain
empirical probability distributions using a
dynamic bucket size,
\eg~\cite{hhh-with-space-savings-arxiv11}. We
leave the adaptation of such algorithms to our
scenario and a quantitative comparison to future
work. 


\subsection{Latency-Aware Layer-4 Load Balancer}
\label{sec:using-response-latency-for-server-load-balancing}

To showcase the utility of real-time response
latency measurement, we present the high-level
design of a layer-4 load balancer that seeks to
optimize average response latency of client
connections. 
%
%
Layer-4 load balancers implement direct server
return (DSR), a mechanism to allow servers to send
responses directly to the client, bypassing the
load balancer~\cite{ananta-sigcomm13} on the path
to the client. Hence, such load balancers exhibit
asymmetric routing.
\OurSystem makes available real-time latency
signals at DSR load balancers through a unilateral
change that does not modify the client, server,
application, or the network.

Below, we describe one way in which a feedback
controller can be designed around response
latency.
We do not claim any novelty in our controller
design; there is a significant body of algorithmic
work on performance-aware request load balancing,
\eg~\cite{c3-nsdi15, prequal-nsdi24,
  nginx-load-balancing, cheetah-nsdi20,
  r2p2-atc19, power-of-2-choices,
  knapsack-conext25, nginx-load-balancing}.
Our design augments a Maglev-hashing-based load
balancer~\cite{maglev-nsdi16}.  Such a load
balancer provides mechanisms to assign weights to
distribute connection load across servers. We use
three key ideas to assign the weights.  First, we
take away weights from servers with latencies
larger than a high watermark, and place them on
servers with latencies smaller than a low
watermark. Second, we limit the addition of weight
to servers that have insufficient recency
(``freshness'') in their latency
measurement. Third, we regress to the mean by
slowly equalizing weights across servers when
latencies have not changed
recently. \App{detailed-lb} presents a full
discussion of these ideas and our
design. \Sec{evaluation} shows the benefits of
real-time latency awareness.

\section{Evaluation}
\label{sec:evaluation}

In this section, we empirically evaluate
\OurSystem by asking the questions below. Our
basic setup is detailed in
\Sec{detailed-experimental-setup}.

\noindent 1. How accurately does \OurSystem
measure response latency for RPC traffic
(\Sec{eval:accuracy}) and Web traffic
(\Sec{eval:chrome}), when compared against
transport-layer RTT?

\noindent 2. What software overheads does
\OurSystem's measurement algorithm impose
(\Sec{eval:overheads})?

\noindent 3. Can real-time awareness of response
latency help drive feedback control? We present a
case study of a latency-optimizing layer-4 load
balancer (\Sec{eval:loadbalancer}).

\noindent 4. How well does each core idea from
\OurSystem work (\Sec{eval:ablation})?




\subsection{Experimental Setup}
\label{sec:detailed-experimental-setup}

This section discusses common aspects of the
experimental setups in our evaluations. We explain
the more detailed setup for each experiment in the
corresponding sub-section.

\Para{Implementation.} \OurSystem runs as a
kernel-bypass software program developed in Linux
eBPF, attaching to the express data path
(XDP~\cite{xdp-conext18}) hook which resides in
the network device driver. For our accuracy
experiments (\Sec{eval:accuracy},
\Sec{eval:chrome}), we use a standalone XDP
forwarder that implements the \OurSystem
algorithm.
The measurement algorithm consists of approximately
500 lines of C eBPF code.

\begin{figure}[]
  \centering
  \[
  \begin{array}{cc}
    \vspace{-15pt}
    \includegraphics[width=0.12\textwidth]{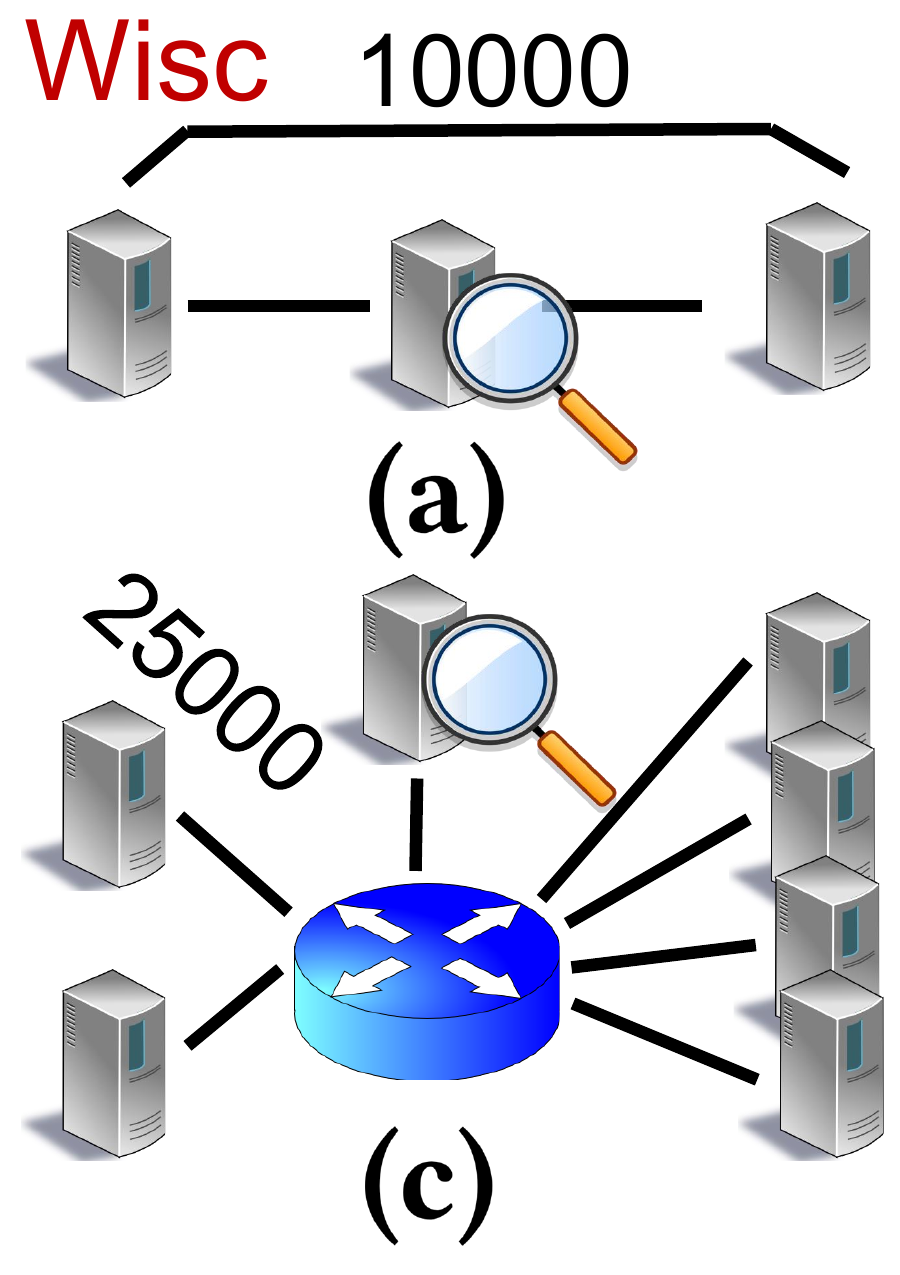} &
    \includegraphics[width=0.32\textwidth]{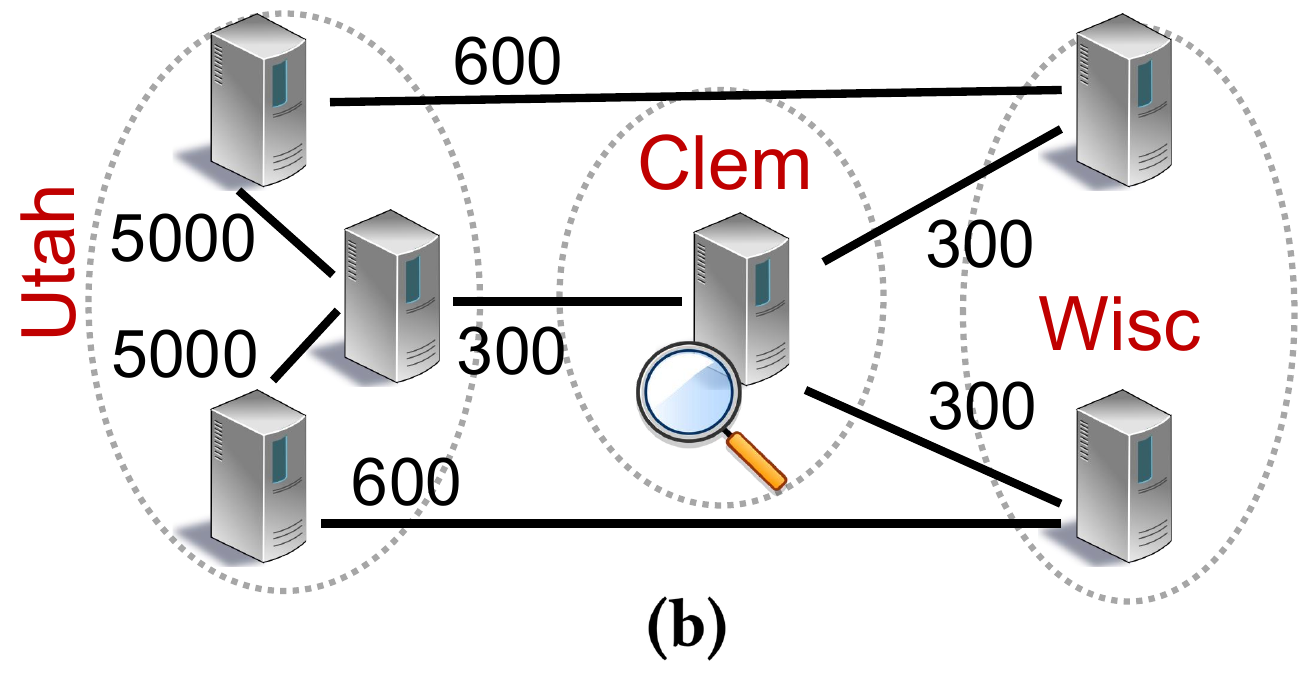} \\
  \end{array}
  \]
  \vspace{-0.4pt}
  \caption{The experimental topologies we use on
    CloudLab~\cite{cloudlab}. Most experiments use
    the single cluster ``triangle'' topology (a).
    We also tested a wide-area topology (b) in
    \Sec{eval:accuracy} and a star topology (c) in
    \Sec{eval:loadbalancer}. Requests flow from
    the clients (left) to servers (right) via the
    measurement vantage point (magnifying
    glass). Responses bypass the vantage
    point. Link bandwidths (Mbit/s) are shown.}
  \label{fig:experiment-topology}
\end{figure}

\Para{Topology.} We set up our implementation of
\oursystem and the client/server workloads on
CloudLab~\cite{cloudlab}.
Unless stated otherwise, our experiments use three
machines in a single cluster, connected in a
triangle topology
(\Fig{experiment-topology}(a)). The three machines
consist of a client, a measurement vantage point,
and a server. Each machine includes two Intel
E5-2630 8-core CPUs, 128GB memory, and a dual-port
10Gb NIC. Requests are routed from the client to
the server via the vantage point.  Responses go
directly from server to client, bypassing the
vantage point. Without network load, the
round-trip time between the client and the server
in this setup is
227 $\mu$s.
%
We also evaluate \oursystem in wide-area settings
with the topology shown in
\Fig{experiment-topology}(b) and on a star
topology shown in
\Fig{experiment-topology}(c). More details on
these are provided in \Sec{eval:accuracy} and
\Sec{eval:loadbalancer}.

\Para{Techniques compared.} We compare \OurSystem
against:


(1) Application-layer response latency ({\bf
  req-to-res}): We instrument the (Web or RPC)
client to compute the time delay between the
transmission of each request and the reception of
the complete response object for that request at
the client's application layer. This is the ground
truth for response latency, labeled {\ct
  req-to-res} on the graphs.

(2) Causal pair delay ({\bf req-to-req}): Even
after receiving a complete response, a client
application may incur additional time to transmit
a \ctp, due to the client thinking
(\Sec{causal-idea}). Since \OurSystem approximates
the time delay between the \cp, for reference, we
show the time between the request and the \ctp at
the client, labeled {\ct req-to-req} on the
graphs.

(3) Transport-layer round trip time (RTT): We
compare \OurSystem against RTT estimators,
specifically (1) {\ct tcptrace} \cite{tcptrace},
an open-source tool that analyzes pcap traces, (2)
{\ct tcp\_probe} \cite{tcp-probe-tracepoint},
which emits the sample RTTs maintained by the
Linux network stack; and (3) syn-ack
delay~\cite{passive-estimation-tcp-rtt-ccr02,
  latency-estimation-networks08}, the time between
the SYN and the ACK packets in the TCP 3-way
handshake. These estimators execute at the client. 
The syn-ack estimator only produces one
measurement per connection.

We ensure the values reported by all of these
estimators are directly comparable by aligning the
transport-layer sequence numbers represented in
the measurements. All estimators except \OurSystem
and syn-ack see traffic in both directions.

\label{sec:DS1}

\label{sec:results}

\Para{Accuracy metrics.} When we compare the
estimate from a technique $T$ against a baseline
$B$, we report:

\vspace{0.1in}

\noindent Absolute error: $latency_B - latency_T$

\vspace{0.1in}

\noindent Relative error:
$\displaystyle\frac{latency_B -
  latency_T}{latency_B}$ (as percentage)

\vspace{0.1in}

Since all latency estimates are non-negative,
relative errors are upper bounded at $100\%$.
Relative errors can take on any negative value,
\ie when $T$ overshoots $B$.

\subsection{Accuracy for RPC Response Latency}
\label{sec:eval:accuracy}

\begin{figure*}[h]
    \centering
    \[
    \begin{array}{cccc}
    \includegraphics[width=0.24\textwidth]{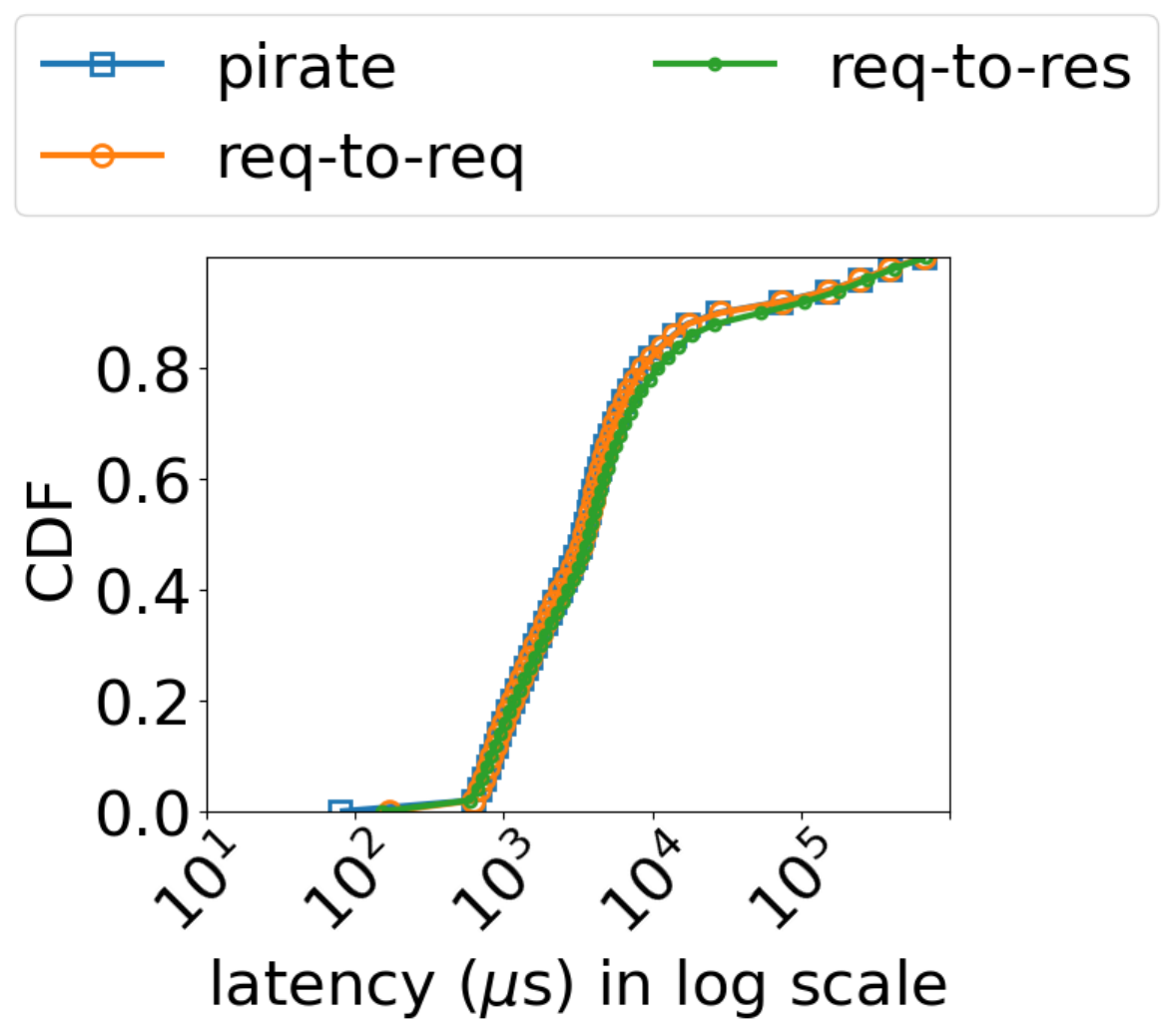} &
    \includegraphics[width=0.23\textwidth]{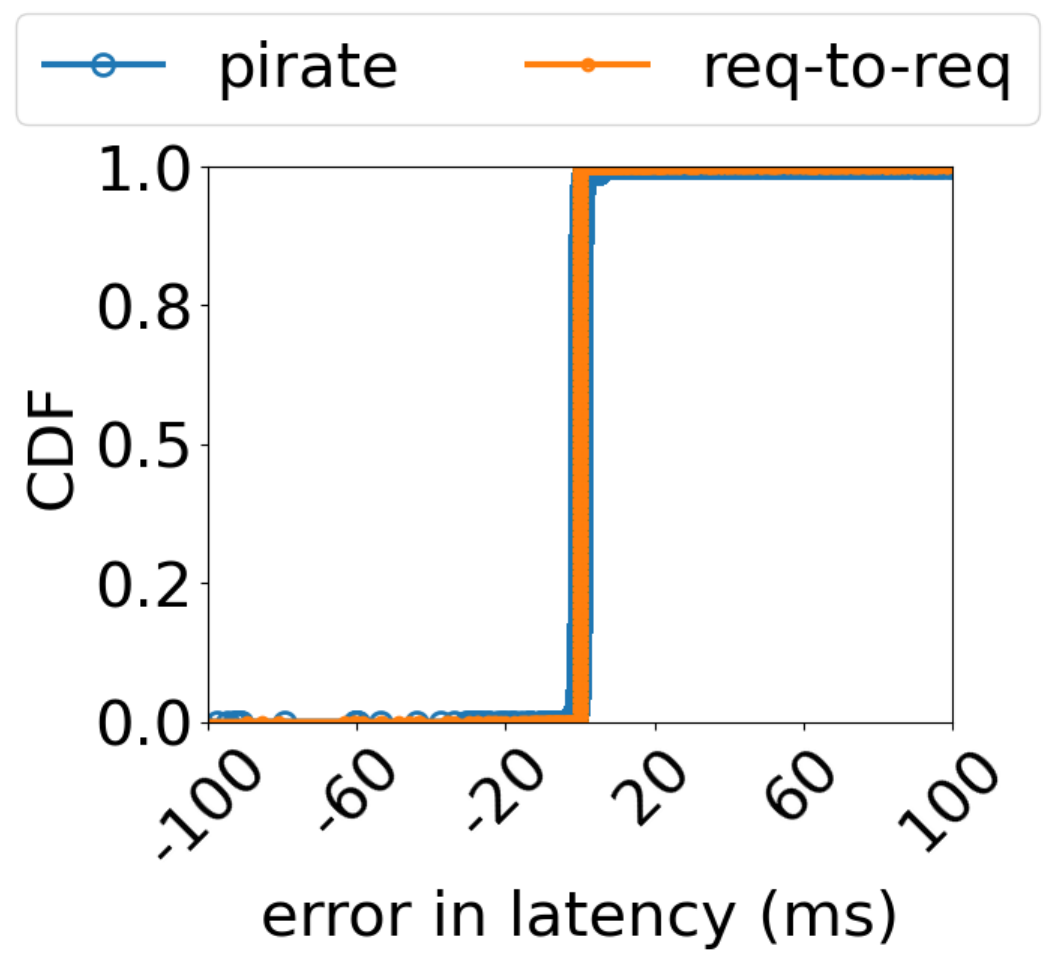} &
    \includegraphics[width=0.24\textwidth]{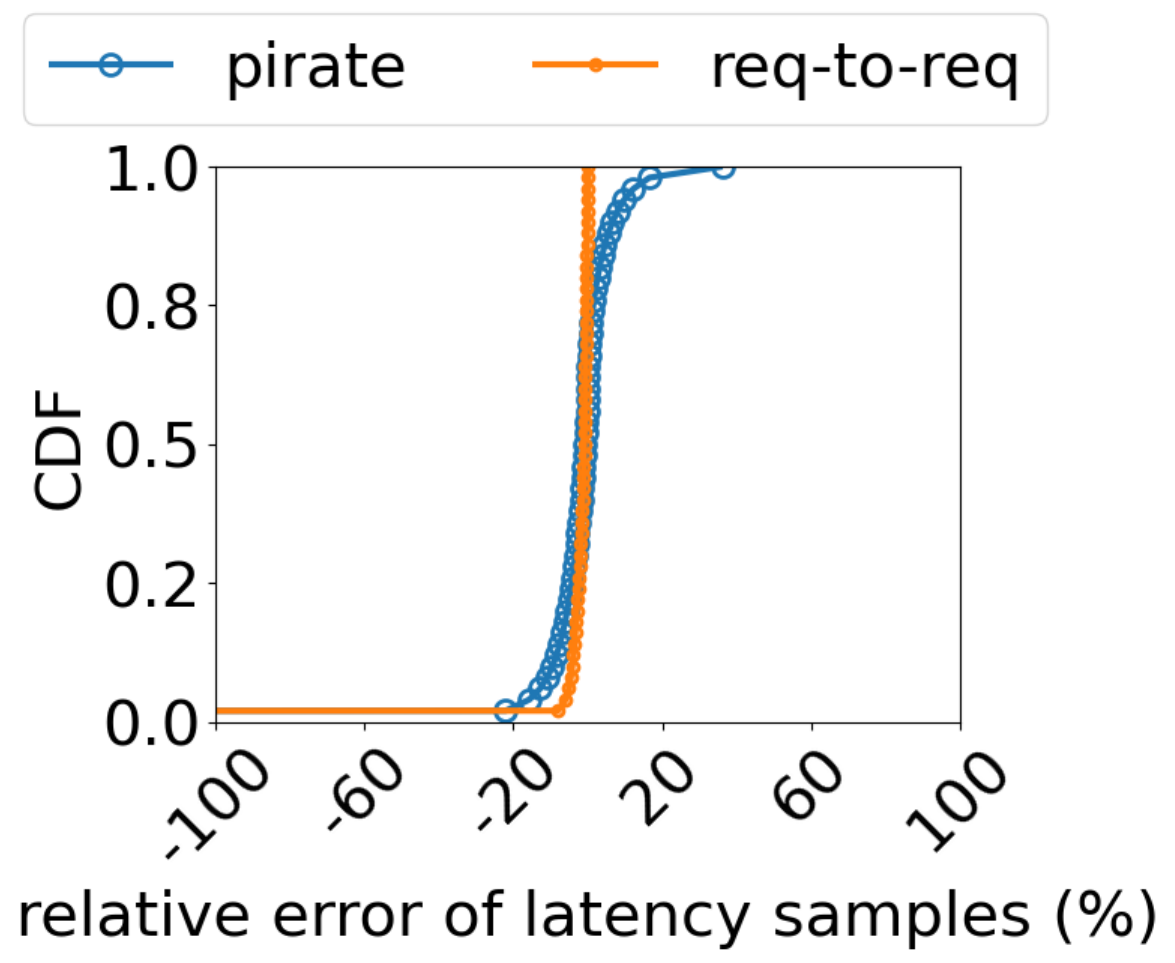} &
    \includegraphics[width=0.23\textwidth]{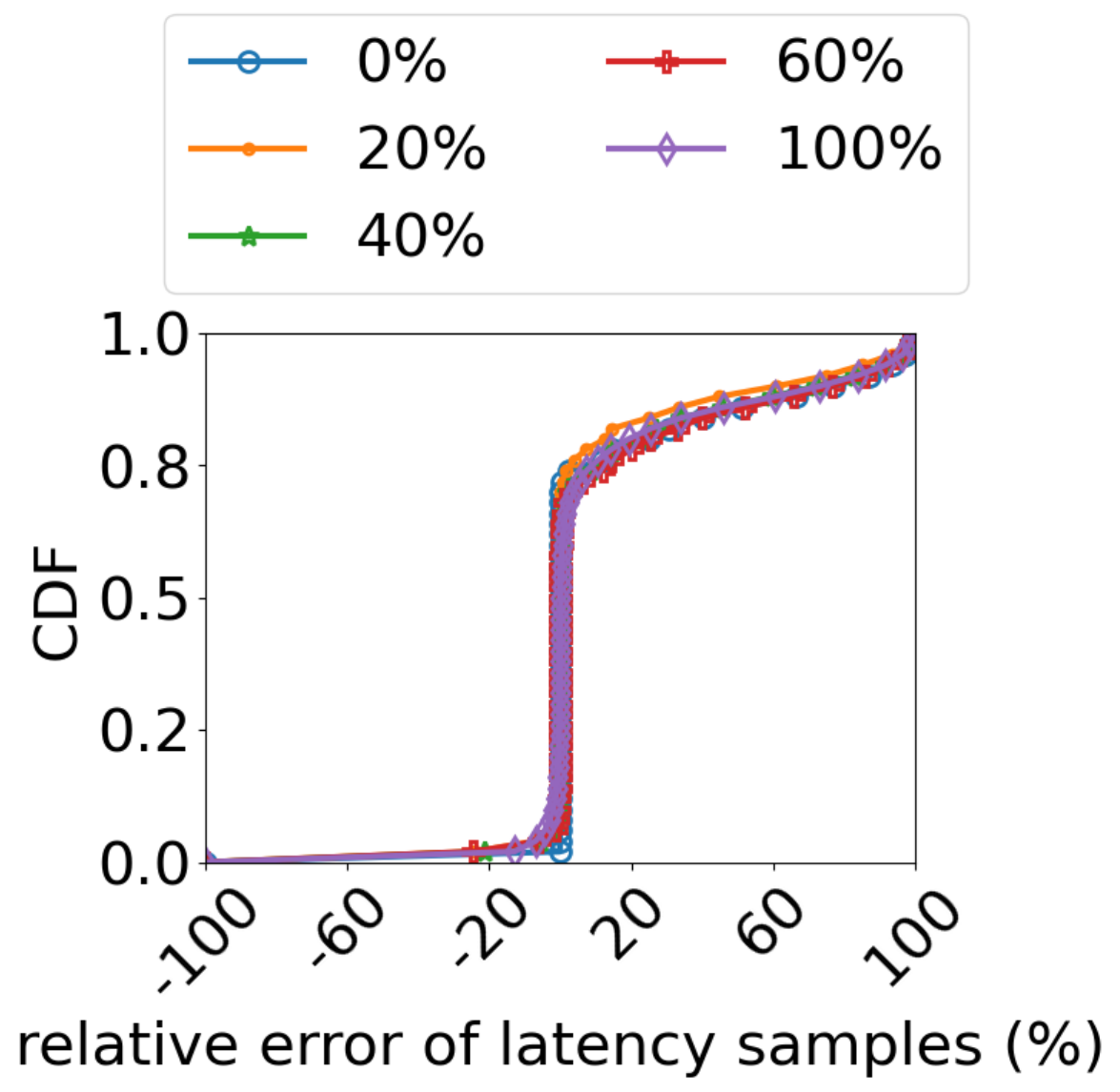}
    \\
    \textrm{(a) All samples (single cluster)} &
    \textrm{(b) Abs. error (single cluster)} &
    \textrm{(c) Rel. error (single cluster)} &
    \textrm{(d) Rel. error (wide area)} \\
    \end{array}
    \]
    \vspace{-0.2in}
    \caption{Accuracy in estimating the response
      latency ({\ct req-to-res}) for RPC
      traffic (\Sec{eval:accuracy}). Subfigures (a)--(c) show
      distributions within a single cluster; (d)
      shows \OurSystem's relative error in a
      wide-area network with cross-traffic offered
      at different fractions of the bottleneck link.
    }
    \label{fig:pipeline}
\end{figure*}

We evaluate how well \OurSystem captures response
latency observed at RPC clients.

\Para{RPC benchmarking framework.} We use the
WebPolygraph suite~\cite{web-polygraph} as our
HTTP/1.1 client and server for
benchmarking. WebPolygraph allows detailed
configuration of workload characteristics such as
(1) the number of benchmarking threads, offered
load (requests per second) initiated from each
benchmarking client thread, and the number of
concurrent requests on persistent pipelined HTTP
connections over TCP; (2) the types of objects
(\eg, {\ct jpeg}, \etc) and their prevalence (\eg,
20\%) among the requested objects; (3) properties
of the dependency tree of objects requested
starting from the root URL requested by the client
(\eg, an object of type $T$ includes $N_a$ objects
of type $a$, $N_b$ objects of type $b$, and so on,
where the values of the $N_{(.)}$ can be drawn
from a custom probability distribution; and (4)
the probability distributions of the sizes of each
object type. The WebPolygraph server is
single-threaded. With our client workload (see
below) and server machine, each client-to-server
connection saturates a single CPU core at a load
of 4K requests/second.

\Para{Server performance variability.} We derive a
time series of CPU allocations for our benchmark
server from a real CPU utilization trace of
Google's Borg cluster jobs~\cite{borg-eurosys15,
  google-cluster-data-2019} (2019 trace). The
trace provides histograms of CPU usage over a
5-minute period, where each sample collected in
the original data is over one second. We select
the time series of the job with the largest
average CPU utilization in the trace, sample from
the histograms, and enforce the resulting CPU
allocation on the WebPolygraph server using Linux
cgroups.
%
%

\Para{Client workload.} The RPC workload we use is
intended to model HTTP API traffic, simulating web
objects fetched from backend servers or objects
fetched for API-oriented
services~\cite{cloudflare-api-traffic-explosion}. We
populate response parameters (types, sizes, object
dependencies) into WebPolygraph by creating a
dataset of objects from the home pages of a random
sample of 20 unique web sites from a top-100
website ranking~\cite{alexa-top-sites}.
Each HTTP request issued by the client fetches one
of 4 kinds of objects (HTML, CSS, JS, image), with
the response size drawn from the empirical size
distribution of the corresponding object type in
our dataset (CSS: median 11.8 KByte, max 1.2
MByte; JS: median 19.6 KByte, max 4.05 MByte;
HTML: median 39.5 KByte, max 2.65 MByte; image:
median 20.8 KByte, max 8.86 MByte).
Unless specified otherwise, our client runs on  single
machine with 20 threads and 200 total connections,
driving an aggregate workload of 2K requests/sec.
To simulate dependent HTTP requests generated by
parsing HTTP responses (\eg, images
embedded in HTML), each response returned by the
server includes references to different kinds of
embedded objects drawn from the corresponding
empirical probability distribution in our dataset.
Connections are persistent and pipelined.
Unless stated otherwise, each WebPolygraph client
connection maintains at most 4 outstanding
requests, comparable to default limits in the HTTP
standard and in popular
browsers~\cite{http-1.1-persistent,
  chrome-6-connection-limit,
  firefox-15-connection-limit,
  raising-firefox-default-connection-value,
  firefox-200-websocket-limit}.
%
%
Each active connection may issue 1--7000
requests, with a median of 2094 HTTP objects
per connection.
%


\Para{Single-cluster setting.}  In a triangle
network topology (\Fig{experiment-topology}(a)),
our client sends a total of 2K requests/second
to the server across all threads and
connections.

\Fig{pipeline}(a) shows the CDF of the response
latencies at the client ({\ct req-to-res}) and the
vantage point (\OurSystem) across all objects over
all connections. We also show the time delay
between the \cp ({\ct req-to-req}) at the client.
%
%
The three distributions are closely aligned,
providing confidence that (i) the time delay
between \cps is a good estimator of response
latency, and (ii) that \OurSystem can estimate the
latter with high accuracy, in realistic settings
where 
the client maintains persistent and pipelined
connections with multiple concurrent
application-layer requests over each connection.
%
%

\Fig{pipeline}(b) shows the CDF of the absolute
error of \oursystem and {\ct req-to-req} against
{\ct req-to-res}, while \Fig{pipeline}(c) shows
the CDF of the relative errors.
The \cp delay is always an overestimate of the
response latency, showing a relative error between
$-10\%$ to $0\%$.
\OurSystem's relative error is less than 1\% at
the median, but stretches beyond $\pm$10\% outside
the $10^{th}-90^{th}$ percentile, and beyond $\pm$15\% at the
tails beyond the $5^{th}-95^{th}$ percentile.
We show an accuracy comparison against
transport-layer RTT estimators later in this
section.

%

\begin{figure}[h]
    \centering
    \[
    \begin{array}{cc}
    \includegraphics[width=0.23\textwidth]{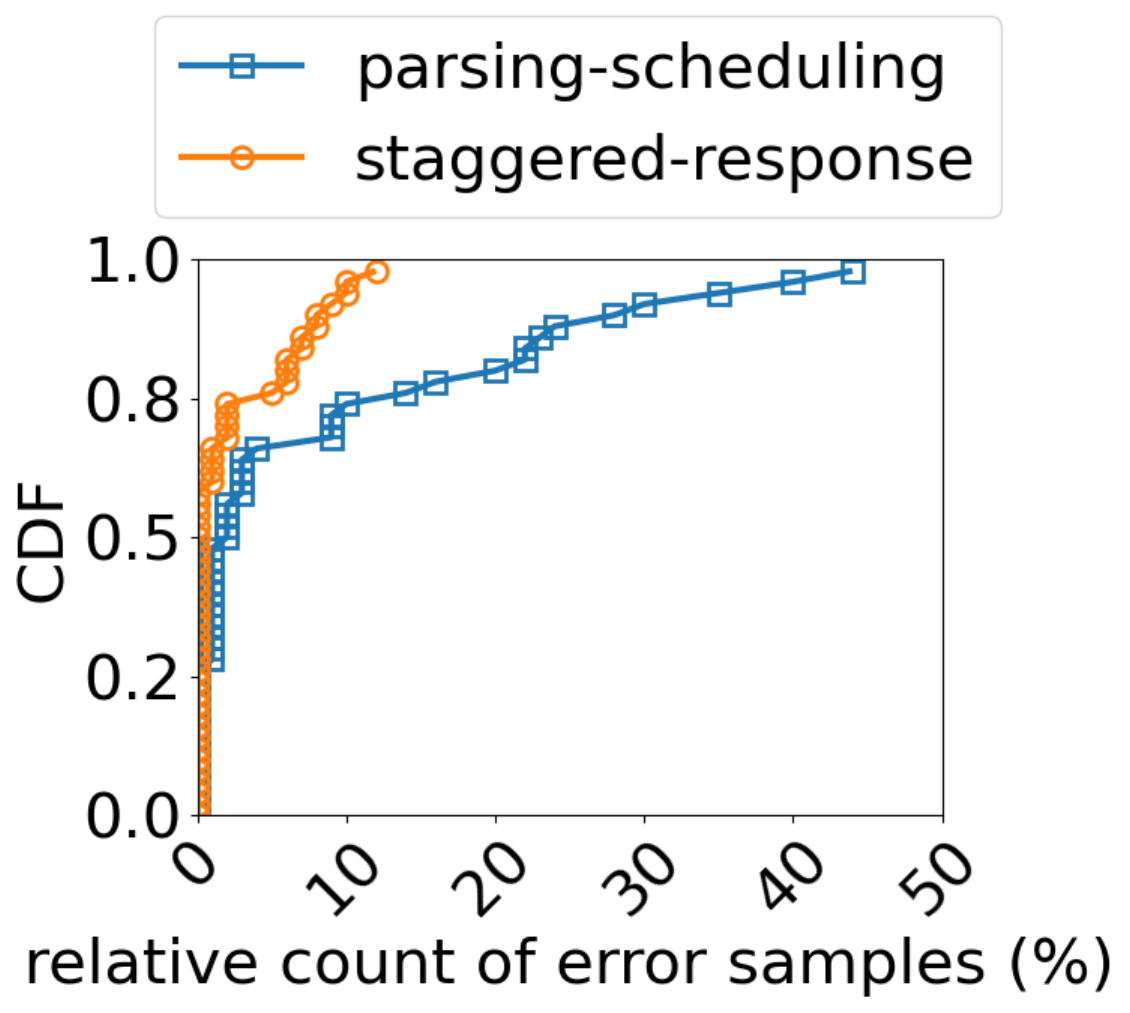} &
    \includegraphics[width=0.23\textwidth]{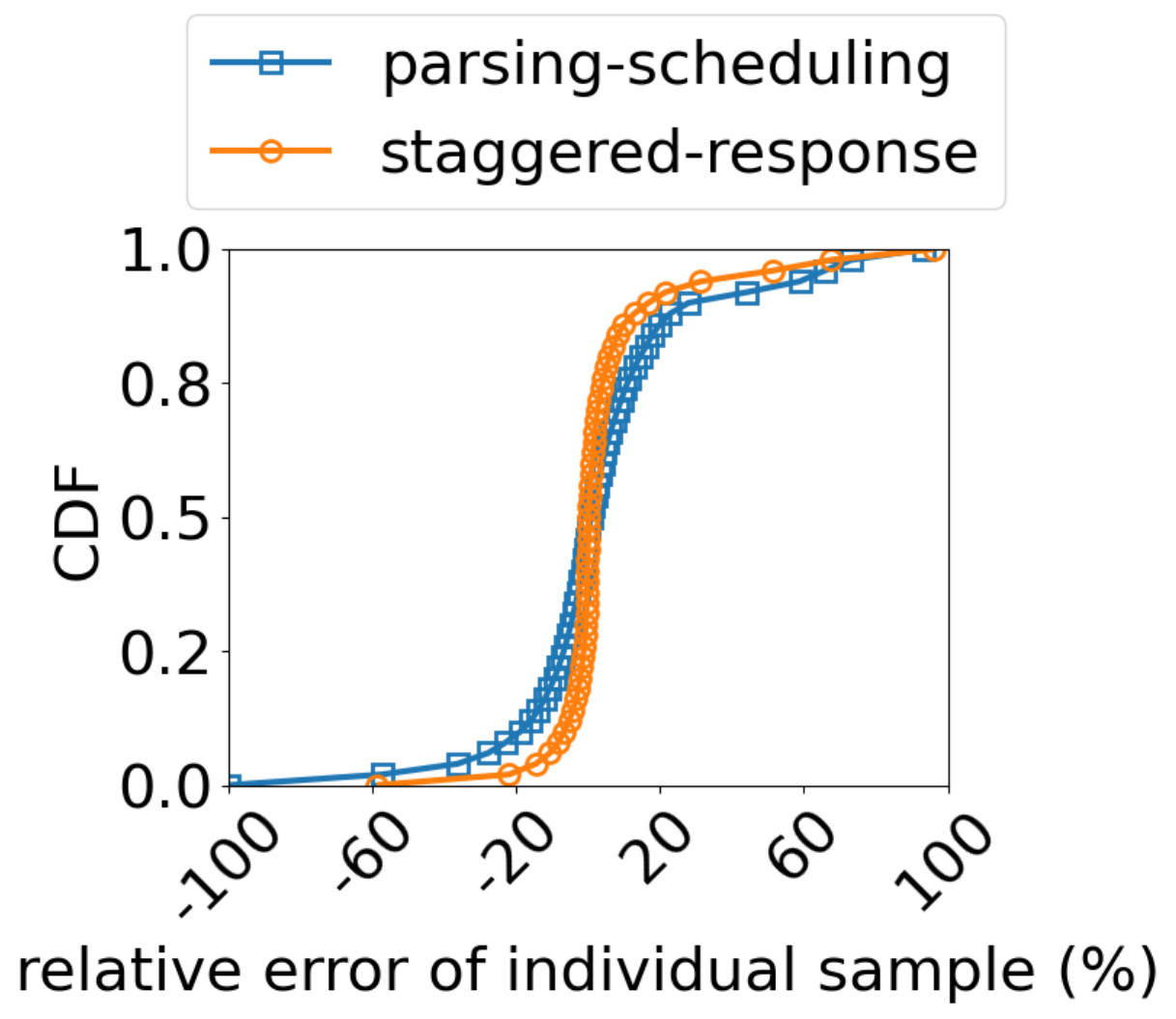}
    \\
    \textrm{(a) Error prevalence} &
    \textrm{(b) Error magnitude} \\
    \end{array}
    \]
    \caption{Classifying \OurSystem's errors in
      estimating response latency, showing
      prevalence (a) and magnitude (b) across
      connections. Discussion in
      \Sec{eval:accuracy} (why does \OurSystem
      make errors?)}
    \label{fig:pipeline_reason}
\end{figure}

\Para{Why does \OurSystem make errors?} An
investigation of when and how significantly
\OurSystem miscomputes response latencies in the
experiment above reveals two kinds of errors.
\Fig{pipeline_reason} measures their prevalence
and magnitude.

The first kind of error, labeled {\ct
  parsing-scheduling}, arises when a client has a
request to transmit, and also has a chance to
transmit (\ie, not subject to flow control or
pipeline concurrency limit), and yet takes
additional time to transmit. This is due to
response processing (\ie, parsing) and delays in
scheduling the client process. \OurSystem
correctly identifies the causal pairs but the
latter delay is not representative of response
latency.

The second kind of error, labeled {\ct
  staggered-response}, occurs when a series of
large response objects (each spanning multiple
network RTTs) causes successive requests to be
transmitted well apart in time. \OurSystem
mistakes the pauses between the causally-unrelated
successive requests to be pauses between
causally-related ones.  Here, \OurSystem
mis-identifies the causal pairs in the first
place.

%
%

\Fig{pipeline_reason}(a) shows the CDF of the
fraction of erroneous samples per connection. The
parsing-scheduling error is more
prevalent. Roughly 70\% of the connections have
fewer than 5\% erroneous samples of either kind.
\Fig{pipeline_reason}(b) shows the CDF of the
average relative error of the erroneous samples
per connection. The magnitude of these errors is
consistent with the overall distribution
(\Fig{pipeline}(c)). Further, the magnitudes of
the two kinds of errors are comparable.

\Para{Wide-area setting.} %
We run the client workload from
\Sec{detailed-experimental-setup} over a wide-area
network spanning three sites
(\Fig{experiment-topology}(b)).
In this topology, a client machine and a
contending traffic source reside in the CloudLab
Utah site and send traffic over a bottlenecked
wide-area link to the Clemson site, which runs the
measurement vantage point. The corresponding
servers reside at the Wisconsin site. Servers
return traffic directly to the client, bypassing
the measurement vantage point. This network
exhibits a propagation round-trip time of 70
milliseconds.  Link bandwidths (Mbit/s) are
labeled in the figure. Different links have
different bandwidths. The bottleneck bandwidths
along the forward and reverse paths differ.

The network includes cross traffic contending with
the measured client and server. The cross traffic
consists of a mix of short- and long-lived TCP
flows whose sizes are sampled from a distribution
measured in a production network at
Microsoft~\cite{dctcp-sigcomm10}.  Flows arrive
and depart throughout the lifetime of the
experiment. The arrival times of new flows are
sampled from a Poisson distribution parameterized
by the average aggregate rate of cross traffic, a
value we set as a fraction of the bottleneck link
capacity of the client-to-server path. We evaluate
different fractions to model different degrees of
bottleneck contention. The bottleneck link on the
request path precedes the measurement vantage
point, and hence the vantage point sees the
effects of link contention on packet timing.
\nop{We do not enforce any explicit server CPU
  allocations in this experiment.}

\Fig{pipeline}(d) shows the relative error of
\oursystem across different degrees of bottleneck
link contention.  \OurSystem's error is close to
0\% for more than 80\% of the measurements across
all cross-traffic configurations.
Comparing \Fig{pipeline}(c) and \Fig{pipeline}(d),
\OurSystem exhibits a higher accuracy in larger
RTT, wide-area settings, than in a single cluster
with a smaller RTT.


\begin{figure*}[ht!]
  \begin{minipage}[b]{0.75\textwidth}
  \centering
  \[
  \begin{array}{ccc}
  \includegraphics[width=0.31\linewidth]{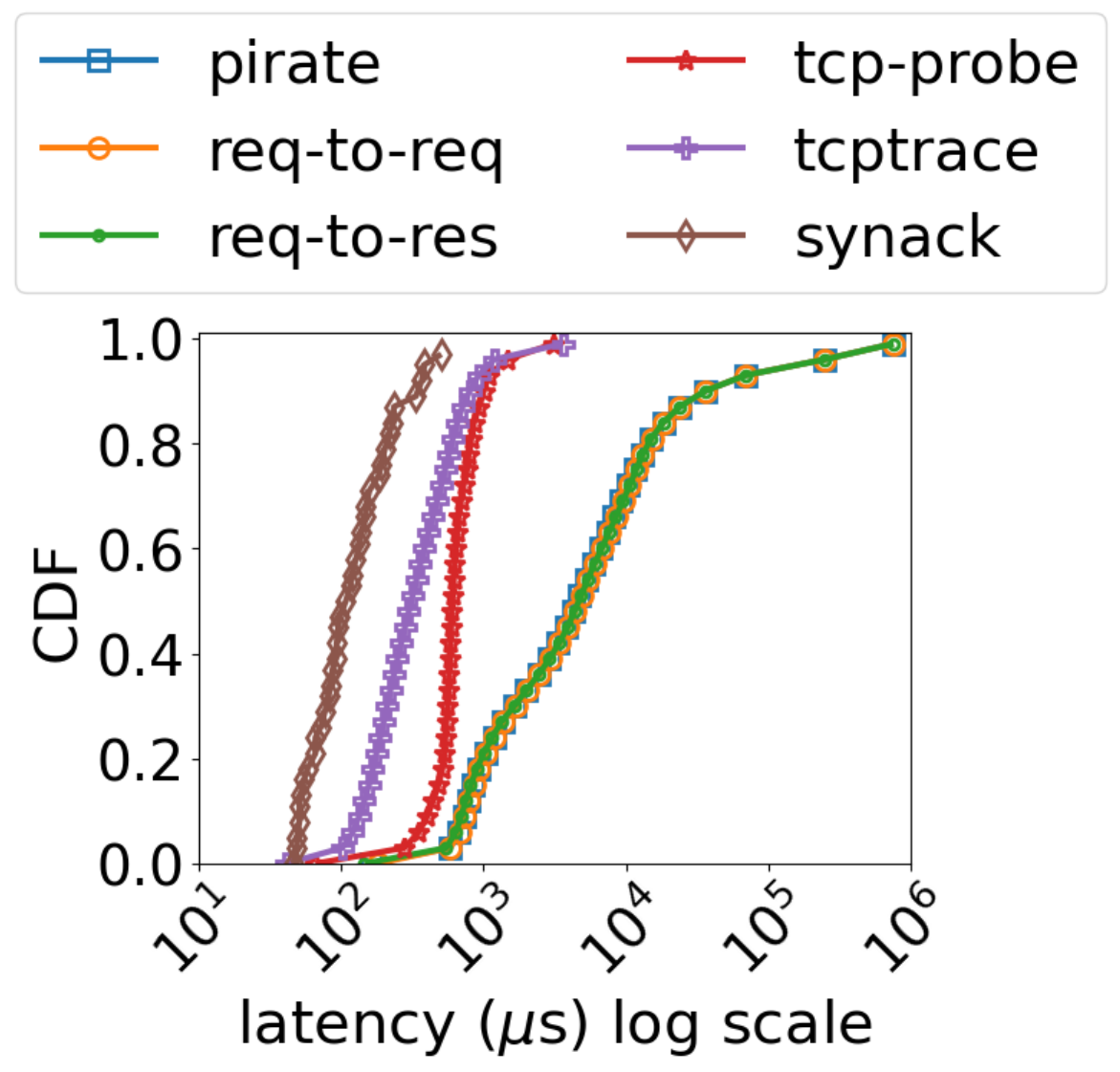} &
  \includegraphics[width=0.31\linewidth]{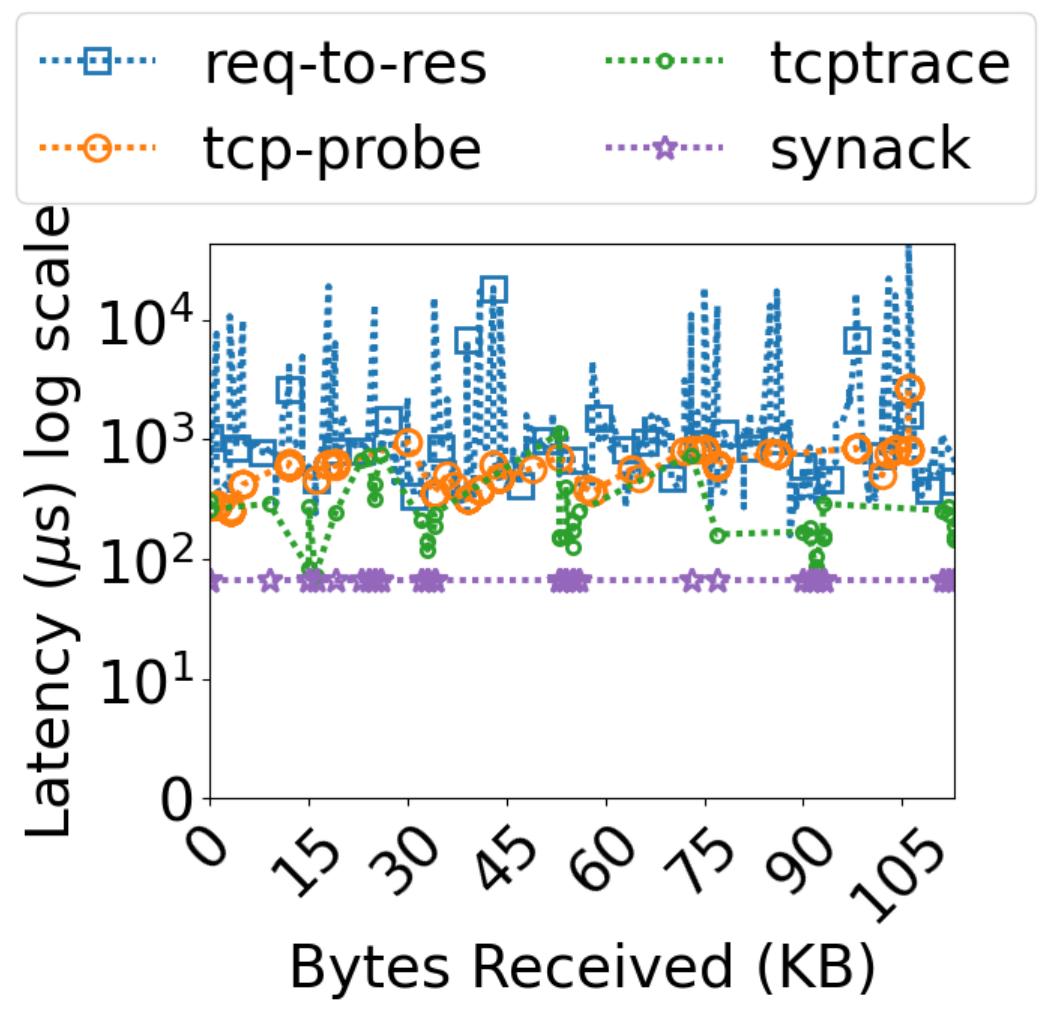} &
  \includegraphics[width=0.31\linewidth]{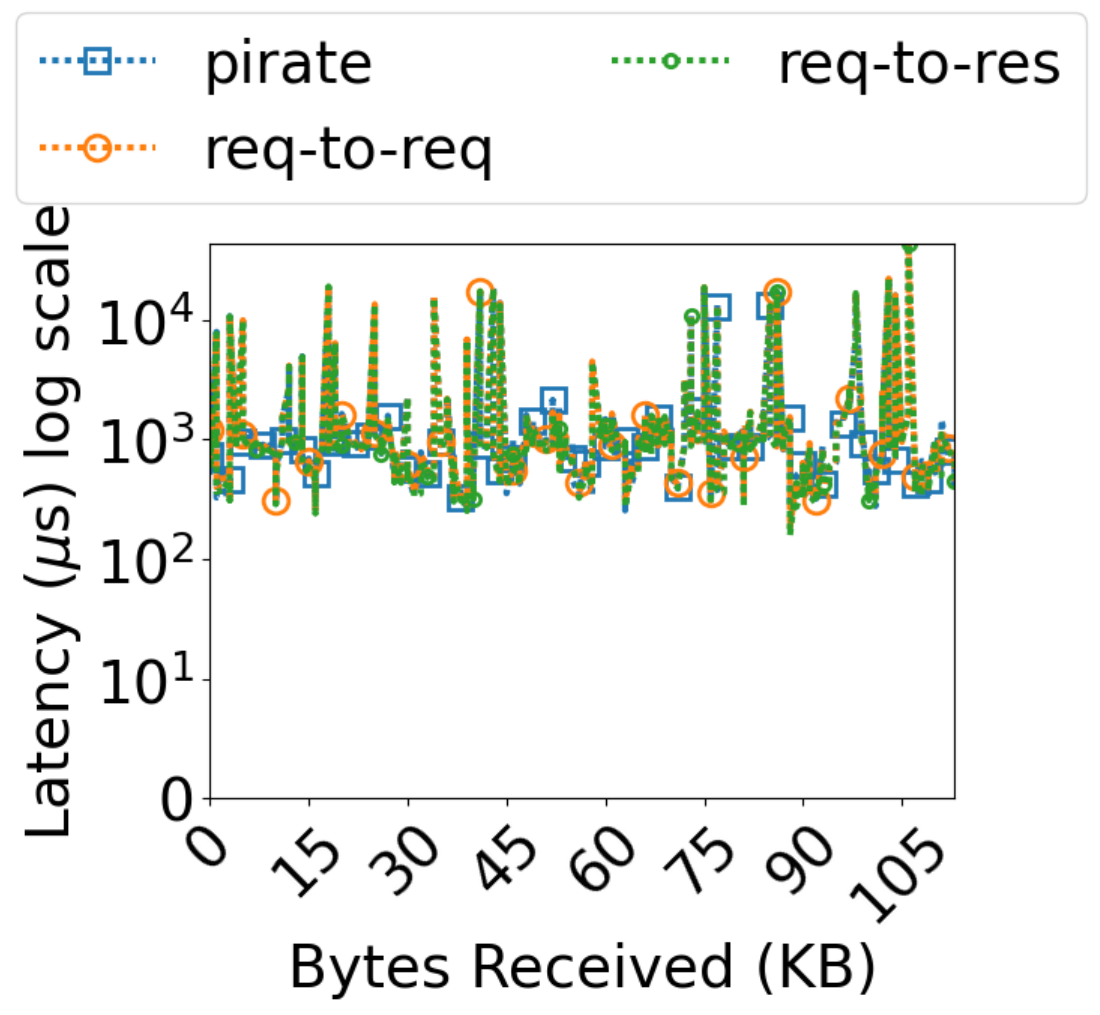} \\ 
  \textrm{(a) CDF of delays} &
  \textrm{(b) Latency vs RTT} &
  \textrm{(c) Latency vs Causal pair delay} \\ 
  \end{array}
  \]
  \caption{ Comparison against transport-layer RTT
    estimators (\Sec{eval:accuracy}): (a) CDF of
    all measurements throughout the lifetime of
    all connections.  (b) and (c): Time evolution
    of the latency of a single (randomly chosen)
    connection during a small window of data
    transfer.  }
  \label{fig:otherSolutions}
  \end{minipage}
  \hfill %
  \begin{minipage}[b]{0.23\textwidth}
  \centering
   \[
   \begin{array}{c}
   \includegraphics[width=\linewidth]{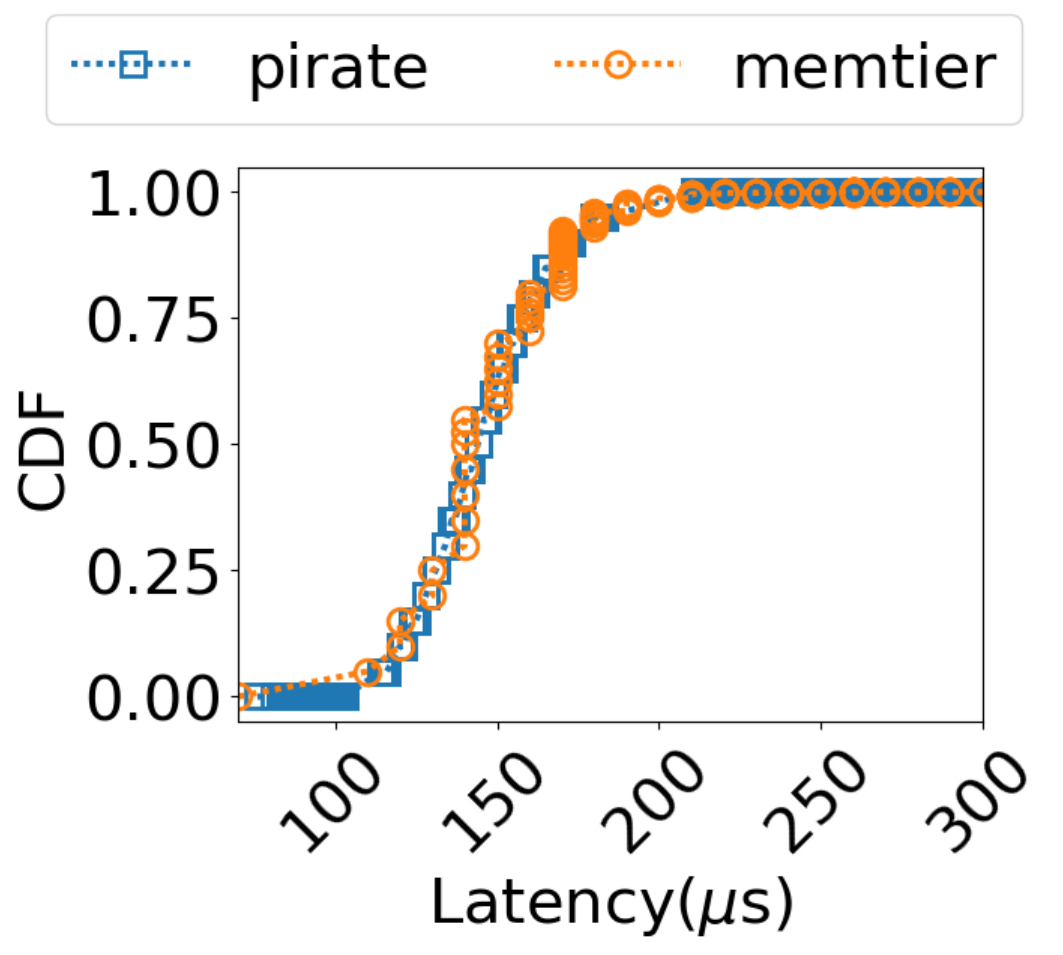} \\
   \textrm{(a) Memtier workload} \\
   \end{array}
   \]
   \caption{Latencies from memtier (ground
      truth) and \oursystem.}
   \label{fig:memtier}
   \end{minipage}
\end{figure*}

\label{sec:eval:comparison-tcp-rtt}
\Para{Comparison against RTT estimators.}
To perform a good-faith comparison of \OurSystem
against transport-layer RTT estimators
(\Sec{response-latency-and-rtt-measurement}), we
run a simplified workload in which the clients
hold only a single request in flight per connection. A
majority of response objects in the workload are
small enough to fit into a single packet, which
increases the likelihood that transport RTTs can
faithfully model response latency. We use the
single cluster setup from
\Fig{experiment-topology}(a).
We compare \OurSystem against {\ct tcp-probe},
{\ct tcptrace}, syn-ack and {\ct req-to-req}
(\Sec{detailed-experimental-setup}). 

%

%
\Fig{otherSolutions}(a) compares the CDF of the
latencies measured by each technique over all
objects across all connections. Even with a
simplified version of a realistic workload, RTT
measures consistently underestimate response
latency, since they are more closely aligned with
the time to the first byte of the response, as
opposed to the last byte. \Fig{otherSolutions}(b)
and (c) show measurements reported over a small
interval of data transfer.  Syn-ack estimation
produces only one measurement per connection, which
does not capture latency variability over
the lifetime of the connection.



\Para{Measuring a memcached RPC client.}
To evaluate the efficacy of \OurSystem over a
different kind of RPC traffic, we run a memcached
benchmarking client, {\ct
  memtier}~\cite{memtier-benchmark} in the
single-cluster topology
(\Fig{experiment-topology}(a)). The client runs
with 10 threads, each with one client connection.
The proportion of GET and SET requests is 1:1. The
response data size is configured to be in the
range 40--10000 bytes.
The experiment is run for 2 minutes.
%
%
\Fig{memtier} shows the CDF of the response
latency measured by the {\ct memtier} application
layer against \OurSystem.  The two
distributions appear highly similar.


\subsection{Accuracy for Web Response Latency}
\label{sec:eval:chrome}

\begin{figure*}[ht!]
  \centering
  \[
  \begin{array}{cccc}
  \includegraphics[width=0.24\textwidth]{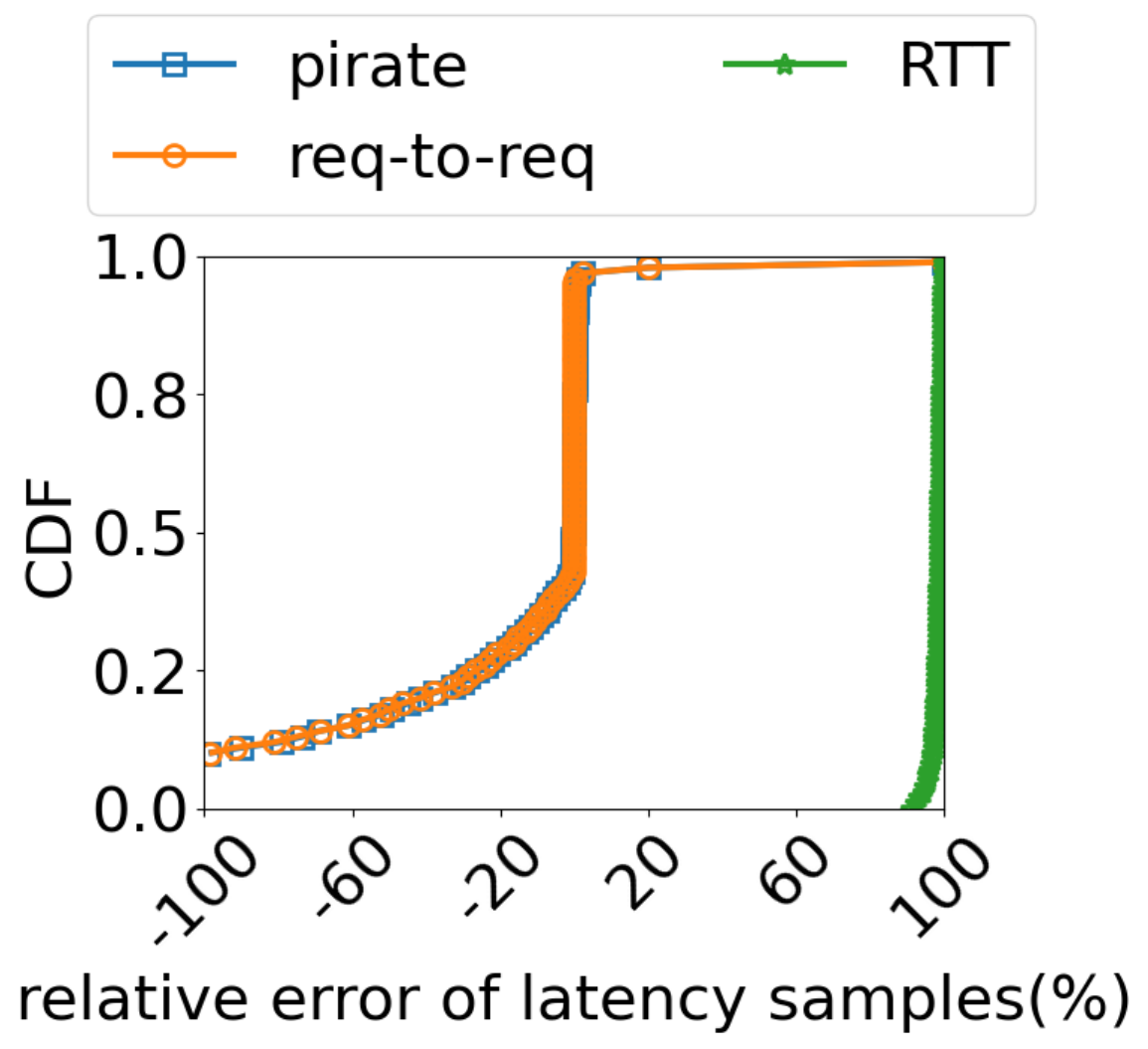} &
  \includegraphics[width=0.22\textwidth]{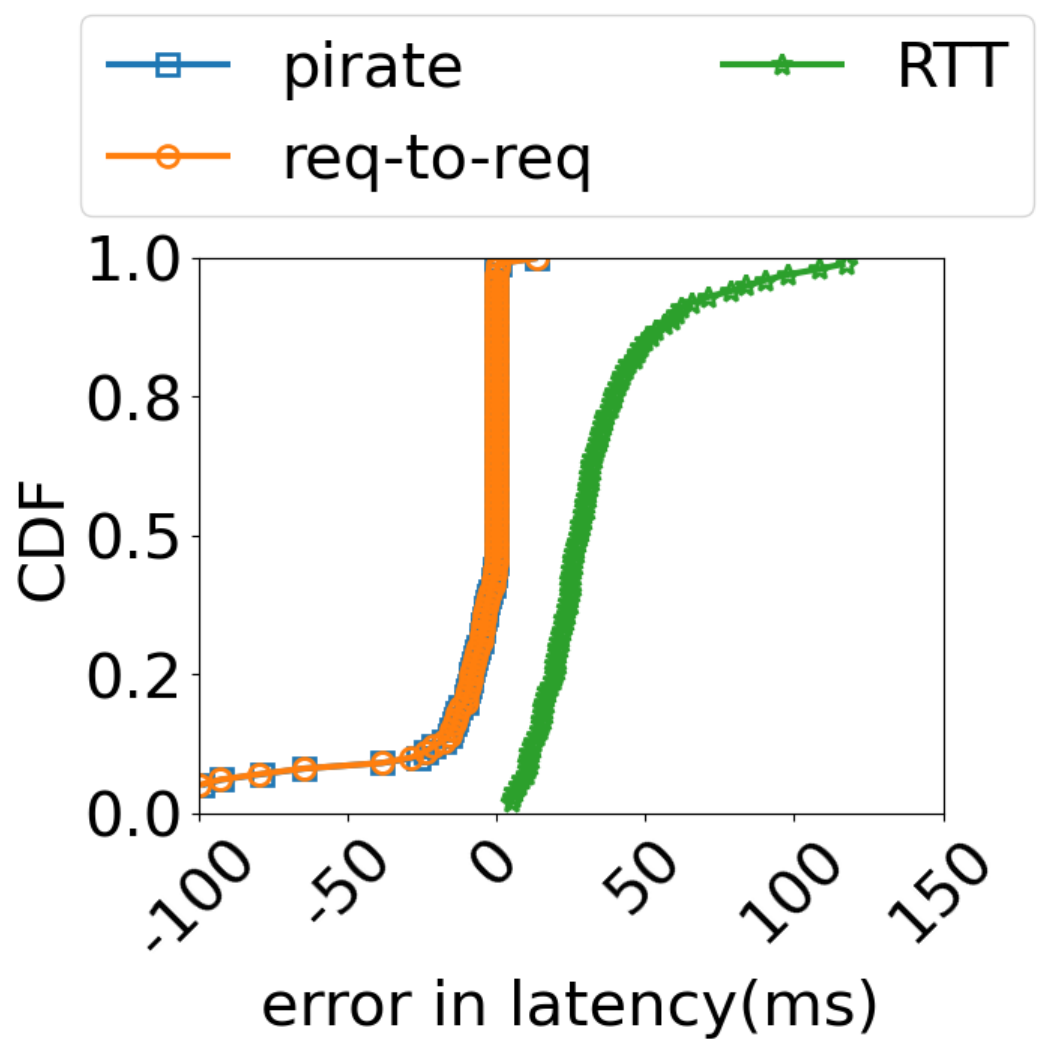} &
  \includegraphics[width=0.24\textwidth]{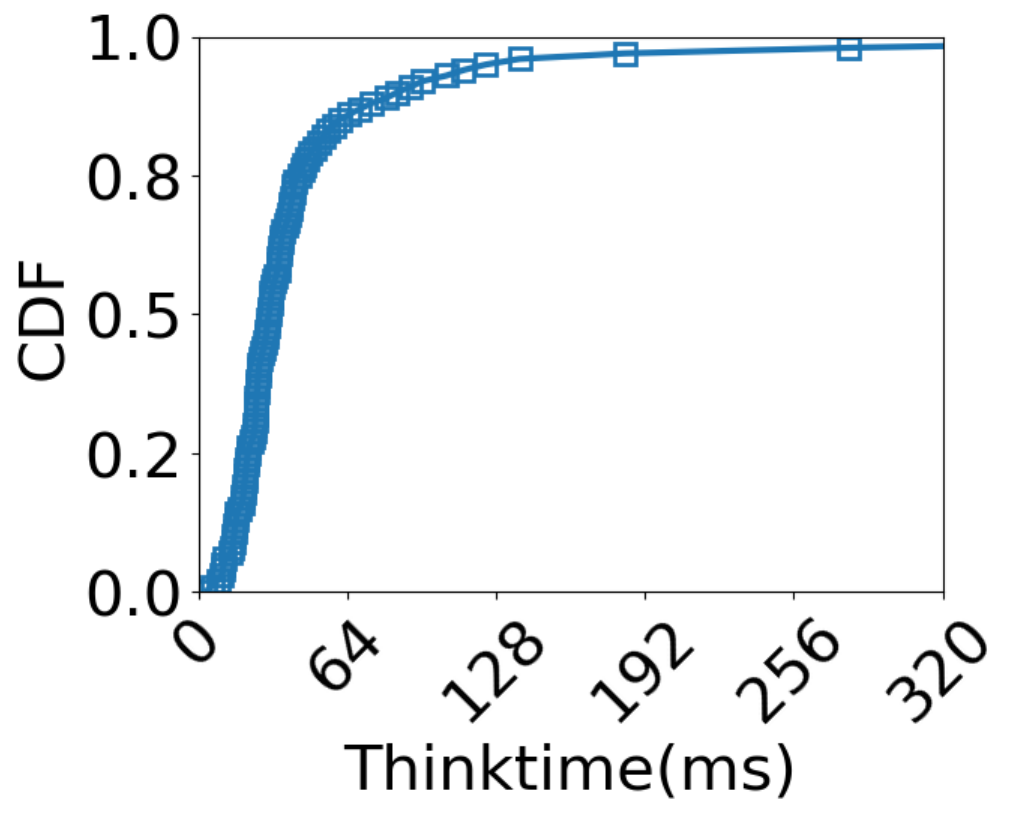} &
  \includegraphics[width=0.24\textwidth]{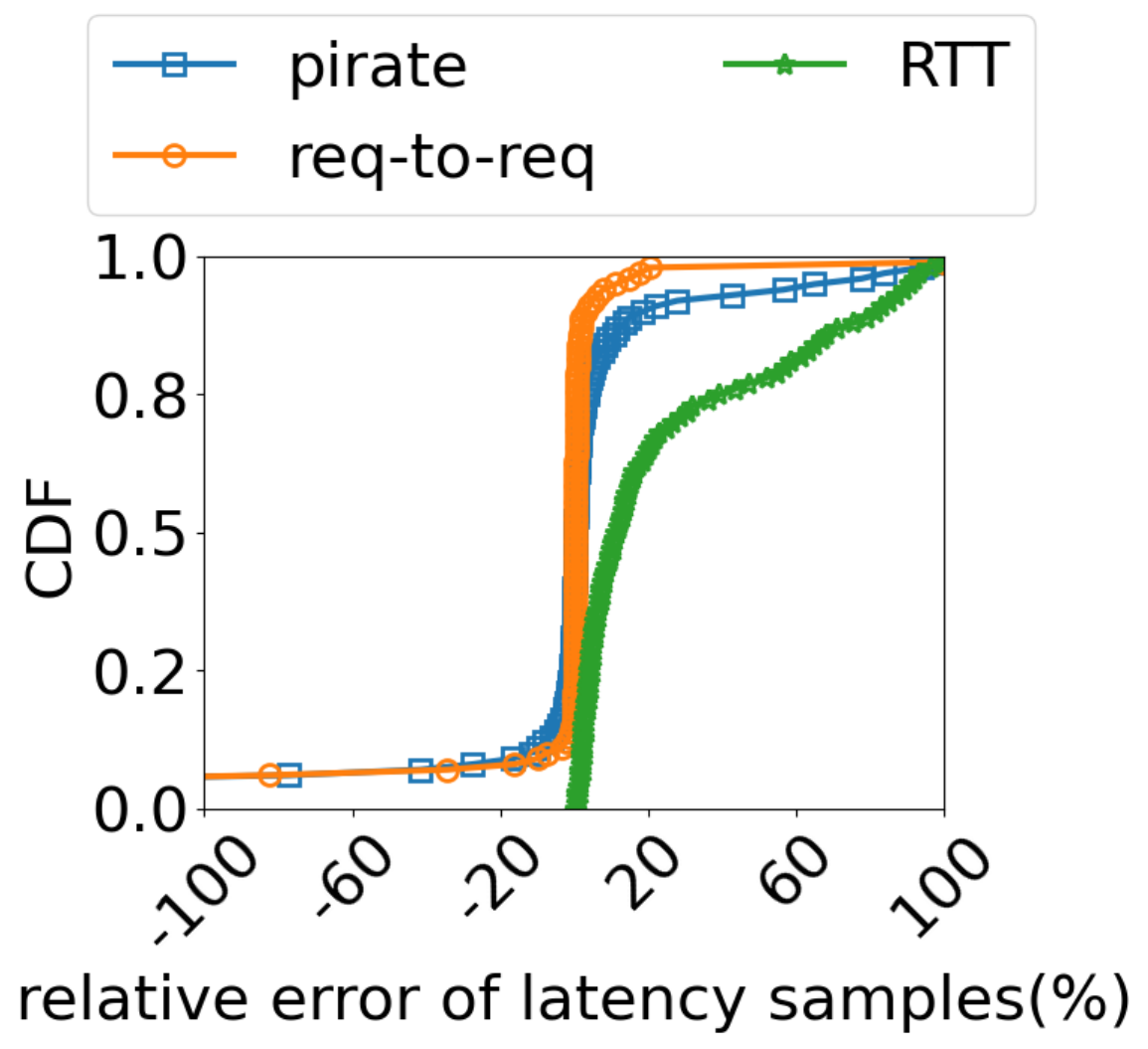}\\
  \textrm{(a) Rel. error} &
  \textrm{(b) Abs. error} &
  \textrm{(c) Think time} &
  \textrm{(d) Rel. error (cellular)} \\
  \end{array}
  \]
  \vspace{-0.2in}
  \caption{ Accuracy in estimating the response
    latency ({\ct req-to-res}) for Web traffic
    (\Sec{eval:chrome}): (a) CDF of relative
    errors; (b) CDF of absolute errors; (c) CDF of
    client think times (\Sec{causal-idea}), the
    primary source of \OurSystem's overestimation;
    (d) CDF of relative errors under an emulated
    cellular network. }
  \label{fig:chrome-evaluation}
  \vspace{-0.1in}
\end{figure*}

We evaluate how well \OurSystem captures response
latency observed at Web clients.
We run experiments using a headless Chrome browser as the Web client, 
driven by a Selenium driver~\cite{selenium-webdriver}. We use
the Nginx server to host a uniform sample of 20 of
the top 100 websites from the Tranco
ranking~\cite{tranco-list}. We run the client and
the server in a single cluster
(\Fig{experiment-topology}(a)). The Web browser
loads the landing page of each site over
HTTP/1.1. The application-layer response latency
({\ct req-to-res}) and \cp delay ({\ct
  req-to-req}) are collected using Chrome
DevTools~\cite{metrics-cdp,
  metrics-network-datareceived,
  metrics-network-responsereceived}.
%
%
The transport-layer RTT is derived from a
client-side packet capture, similar to {\ct
  tcptrace}.

\Fig{chrome-evaluation}(a) and (b) show the
relative and absolute errors of \OurSystem, \cp
delay, and transport-layer RTT, with the response
latency as the baseline. \OurSystem closely agrees
with the \cp delay. Even though the application
exhibits \cps, the \cp delay deviates from the
response latency due to noticeable client think
times
(\Sec{causal-idea}). \Fig{chrome-evaluation}(c)
shows the CDF of the client think time, \ie the
time between when the application layer receives a
response and transmits its corresponding \ctp.
\OurSystem captures the response latency to within
20\% roughly 80\% of the time. In comparison,
using the transport-layer RTT underestimates
application-layer response latency by over 90\%
more than 95\% of the time.

%
%

\nop{

Furthermore, \Fig{chrome-evaluation}(c) quantifies
the queuing delay and think time experienced by
each resource, from the moment the renderer
submits the request until the network transmits
the request onto the wire.

{\color{purple} To evaluate the efficacy of
  \OurSystem using a real browser and a real
  website, we ran experiments with a Selenium
  headless Chrome browser on the client side and
  Nginx hosting 20 selected sites from the top 100
  of the Tranco list of 1 million sites on the
  server side. The workload was run on the single
  cluster topology, illustrated in
  \Fig{experiment-topology}(a). The client loads
  the initial webpage of the 20 selected sites
  over HTTP 1.1. We collected groundtruth
  measurements from CDP events of the
  browser}

{{\color{brown}, (Note the figure with
    definition is there in repository under the
    name CDP\_timediagram.pdf)definition of which
    is presented in
    \Fig{chrome-evaluation}(a)}}

  {\color{purple}. Additionally,
    we gathered the round-trip time from the
    client's packet
    capture. \Fig{chrome-evaluation}(a) and (b)
    demonstrate the relative and absolute error of
    \OurSystem relative to the baseline. The
    baseline used is the request-to-response
    calculated from CDP events at the
    client. Notably, \OurSystem is in close
    agreement with request-to-request observed at
    Chrome's network service. Even though the
    workload exhibit causal pair properties this
    experiment demonstrates the impact of think
    time on \OurSystem's approximation, positive
    side of error in \Fig{chrome-evaluation}(a) is
    due to renderer's processing delay of response
    while network service send out any pending
    request in the queue and negative side of
    error is due to think time. Negative errors
    are capped at -100\%. Furthermore,
    \Fig{chrome-evaluation}(c) quantifies the
    queuing delay and think time experienced by
    each resource, from the moment the renderer
    submits the request until the network
    transmits the request onto the wire. }

}
  
\Para{Cellular network setting.} We run
our Web browser client over a cellular link
emulated with mahimahi~\cite{mahimahi-atc15}, with
an added delay of 20 ms and the time-varying
capacity of a U.S. cellular network
link~\cite{mahimahi-verizon-trace}. \Fig{chrome-evaluation}(d)
presents the errors relative to the response
latency {\ct req-to-res}. \OurSystem incurs errors
within $\pm$20\% about 80\% of the time. In
comparison, the median error of the
transport-layer RTT (which always underestimates
the response latency) is close to 20\%.
  
\label{sec:eval:cellular}

\nop{

{\color{purple} To evaluate the efficacy of
  \OurSystem under cellular environment we repeat
  the browser experiments in \Sec{eval:chrome} in
  a cellular environment setup using mahimahi. We
  run the Selenium headless browser within a shell
  that simulates 20ms delay and time-varying
  capacity of U.S. cellular networks as
  experienced by a mobile user with Verizon
  traces. We present the results for this
  experiment in \Fig{chrome-evaluation}(d). The
  baseline used is the request-to-response
  calculated from CDP events at the client.}

}

\subsection{Software Overheads}
\label{sec:eval:overheads}

\begin{figure*}[ht]
    \centering
    \[
    \begin{array}{cccc}
    \includegraphics[width=0.32\textwidth]{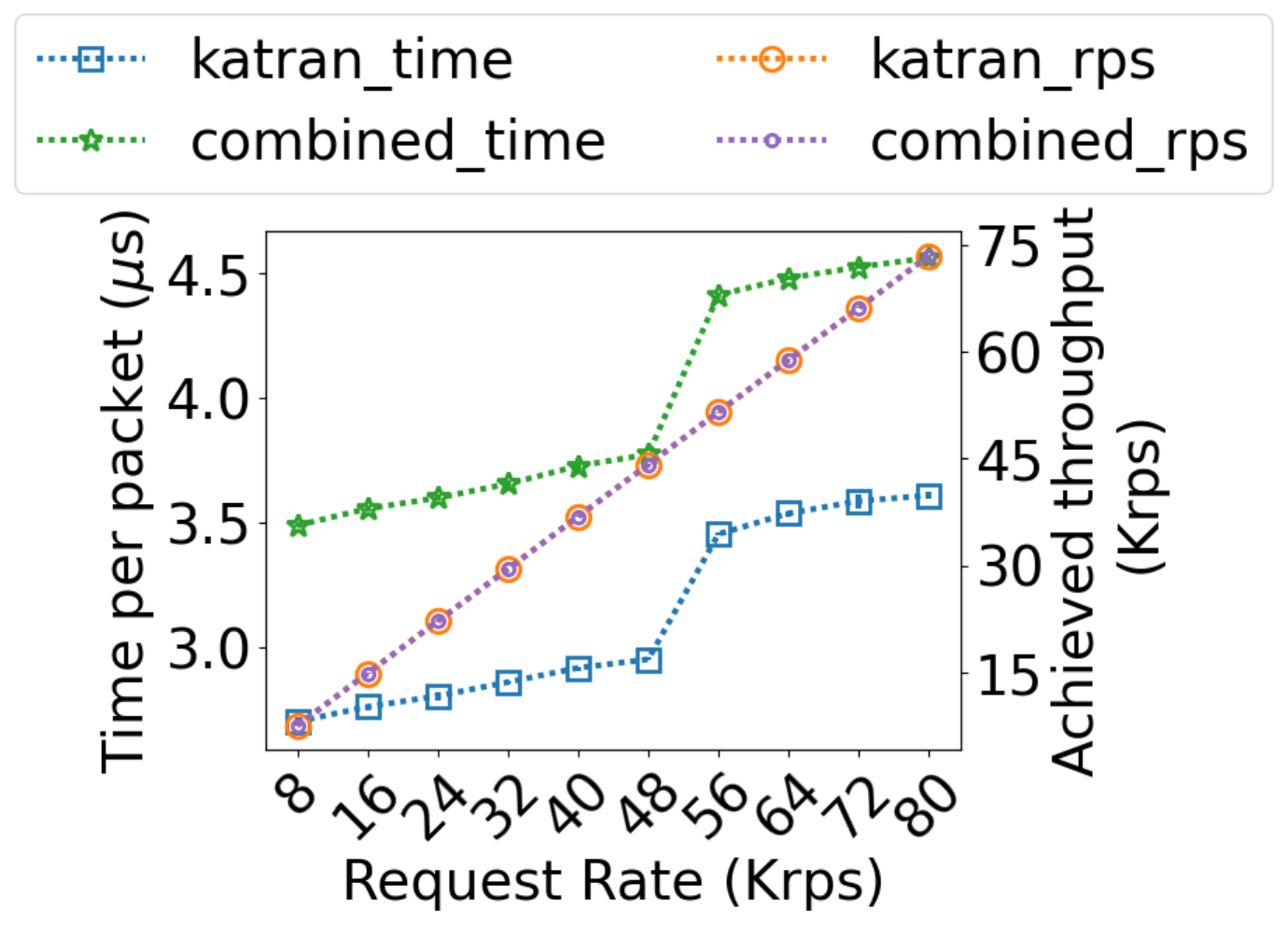} &
    \includegraphics[width=0.32\textwidth]{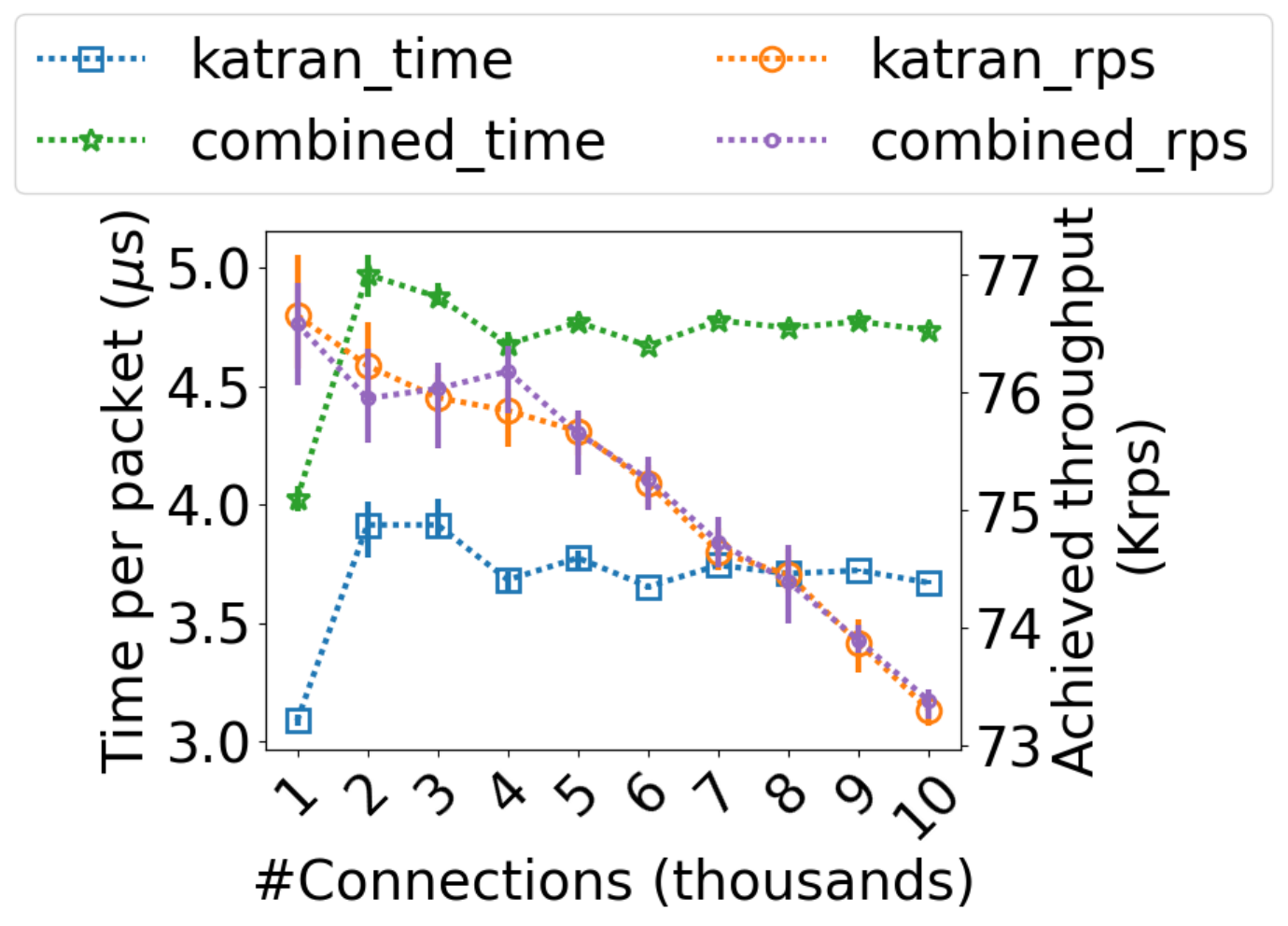} &
    \includegraphics[width=0.30\textwidth]{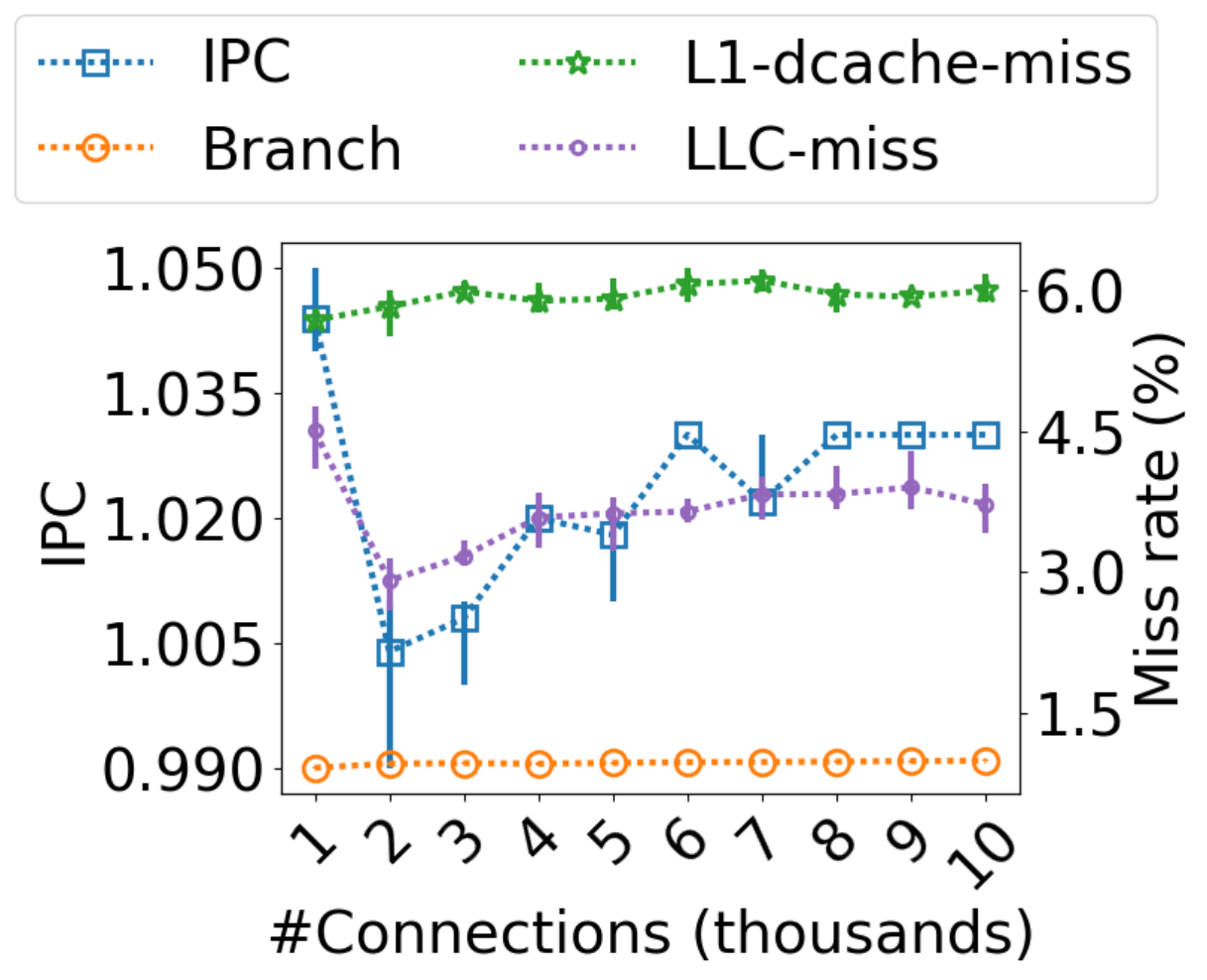}
    \\
    \textrm{(a) RPS \& Time vs load} &
    \textrm{(b) RPS \& Time vs \# conns.} &
    \textrm{(c) {\ct perf} vs \# conns. (combined)} \\
    \end{array}
    \]
    \vspace{-0.2in}
    \caption{The CPU and latency overheads of
      \OurSystem in the context of a layer-4 load
      balancer. See discussion in
      \Sec{eval:overheads}.
    }
    \label{fig:overhead-cpu-latency}
\end{figure*}


We evaluate the CPU overheads of the \OurSystem
algorithm and its XDP implementation in the
context of another XDP packet-processing
application, Katran~\cite{katran-description}, a
layer-4 load balancer used in production clusters
at Meta. We use the single-cluster setup
(\Fig{experiment-topology}(a)). The vantage point
serves as the device under test. 
We compare the performance of (solo) Katran with a
combined system running \OurSystem and Katran
chained in series via an eBPF tail
call~\cite{ebpf-tail-call}.
%
%
The client runs {\ct wrk2}, a web benchmarking
client, with 12 threads and a configurable offered
load (requests/sec) and connection count.
We measure the packet-processing throughput,
latency, and CPU utilization. We present the
average across 5 runs with error bars representing the
minimum and maximum values across the runs. Each
performance test runs for 30 seconds. We also
measure CPU microarchitectural performance
counters using {\ct perf}~\cite{perf, intel-pcm,
  bpf-profiling} throughout the tests, reporting 
only the contributions from the XDP software.

We have evaluated the combined system and solo
Katran up to an offered load of 80,000
requests/sec and an active connection count of
10,000.
%
These limits have been chosen to be competitive
with some publicly known high-performance
middlebox
deployments~\cite{digital-ocean-lb-deployment}.

\Fig{overhead-cpu-latency} shows the results.
\Fig{overhead-cpu-latency}(a) shows the per-packet
latency and request-processing throughput of the
two systems compared, as we vary the offered load
in requests/sec. The number of active connections
is fixed at 10,000. The XDP programs execute on all
16 CPU cores
(\Sec{detailed-experimental-setup}). Integrating
\OurSystem into Katran adds an overhead of 800 ns
on average compared to the runtime of solo Katran,
while achieving throughput comparable to that of
solo Katran.

In \Fig{overhead-cpu-latency}(b) and (c), we fix
the total offered load at 80,000 requests/sec and
vary the number of active client
connections. Subfigure (b) shows that \OurSystem
offers stable packet-processing times and
throughput upto 10,000
connections. Microarchitectural metrics reported
in subfigure (c) show that instructions per cycle,
branch miss rate, L1 data cache miss rate, and LLC
miss rate of the combined system are stable under
varying connection counts. The combined system
exhibits similar metrics as Katran (curves
elided).

Separately, we measure the CPU utilization of the
combined system when fixing the number of active
connections at 10,000, and varying the offered
load (requests/sec) up to 80,000. For simplicity
of interpretation, we use Linux Receive Side
Scaling (RSS~\cite{rss}) to redirect all incoming
requests to a single fixed CPU core. 
\OurSystem's CPU contribution is a single-digit
percentage across all offered loads, and it is
lower than that of Katran ($\sim$20\%) and the CPU
idle thread ($\sim$75\%).

\nop{
{\color{purple}In \Fig{overhead-cpu-latency}(a) and (b), we fix
the number of connections to 10K and vary the
offered load (Krequests/sec). Subfigure (a)
reports the CPU utilization measured using {\ct
  perf} events~\cite{perf}, in comparison to
Katran, tail called by \OurSystem, and the CPU idle thread. \OurSystem's CPU usage
is relatively lower than that of Katran. For Subfigures (b), (c), and (d)
the Katran curve represents results when Katran is run without \OurSystem and
the \OurSystem curve represents results when \OurSystem is integrated with Katran, i.e., 
\OurSystem makes a tail call to the Katran load balancer program. 
\Fig{overhead-cpu-latency}(b) compares the per-packet processing time and achieved throughput of 
\OurSystem with Katran. \OurSystem adds an overhead of 800ns on average compared to runtime of Katran. 
\OurSystem achieves comparable throughput to that of Katran.}

{\color{purple}In \Fig{overhead-cpu-latency}(c) and (d), we fix
the total offered load at 80K requests/sec and
vary the number of active client
connections. Subfigure (c) shows that \OurSystem
offers stable packet-processing times and
throughput upto 10K connections. An investigation using system
performance counters~\cite{perf}, reported in
subfigure (d), shows that the branch miss
and data cache miss rates of \OurSystem are comparable to Katran.}

}


XDP requires provisioning memory in advance for
lookup data structures. By default, Katran
provisions a hash table for 64K connections,
incurring 458 MBytes. In comparison, supporting
the same number of connections in \OurSystem
requires 12 MBytes (154 bytes/connection).

\begin{figure*}[t]
    \begin{minipage}[b]{0.75\textwidth}
    \centering
    \[
    \begin{array}{ccc}
    \includegraphics[width=0.32\linewidth]{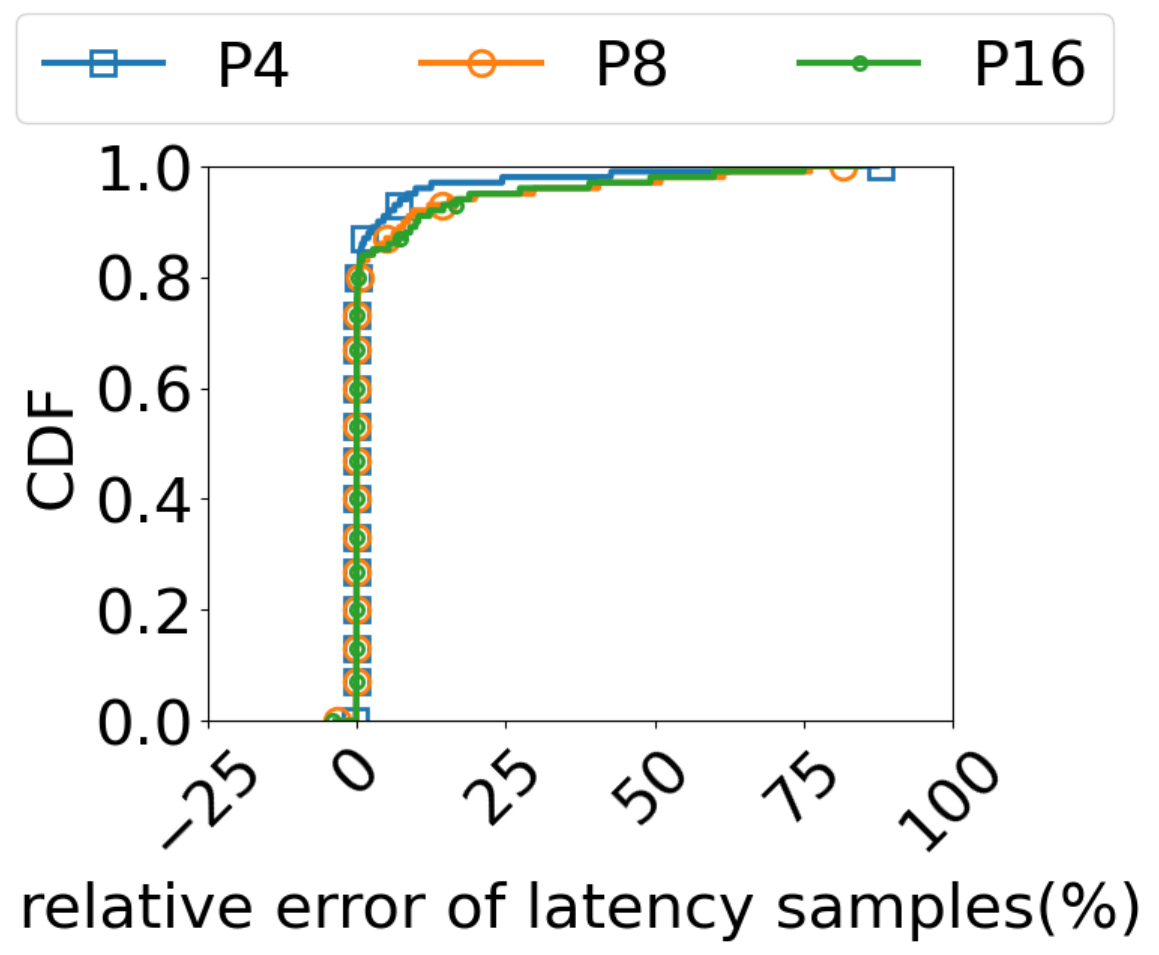}&
    \includegraphics[width=0.31\linewidth]{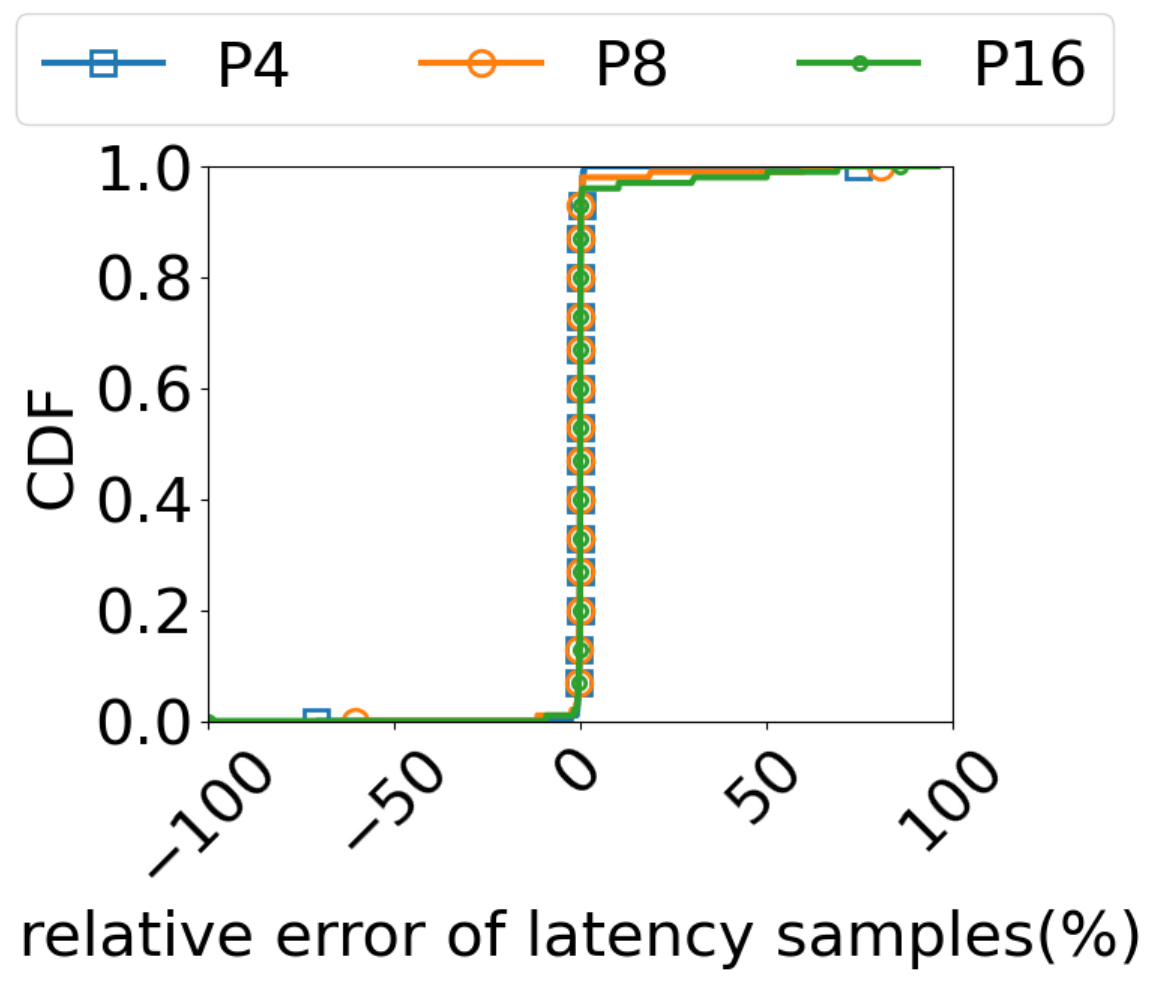}&
    \includegraphics[width=0.31\linewidth]{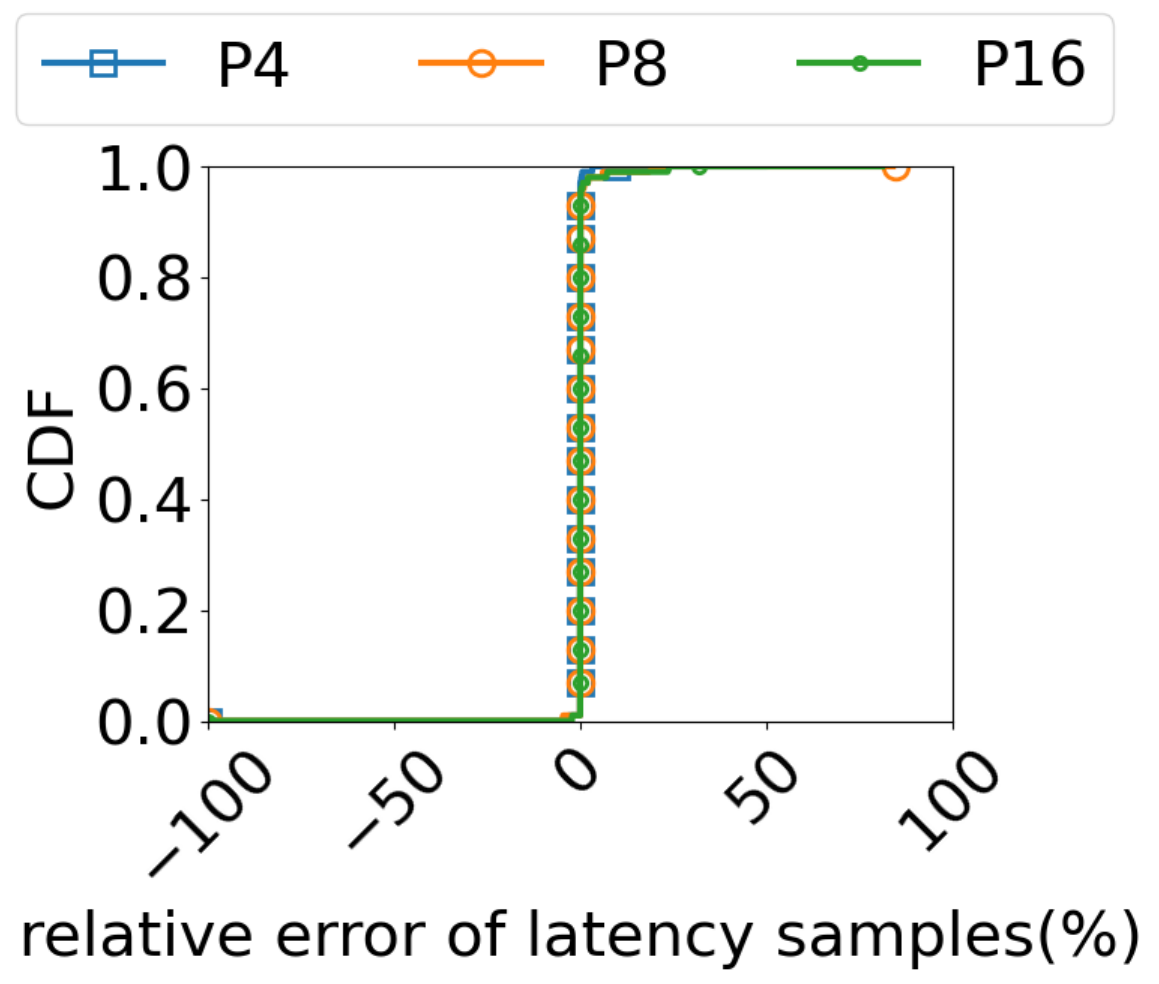}\\
    \textrm{(a) \cp \vs resp. latency } &
    \textrm{(b) prom. gap \vs \cp } &
    \textrm{(c) \OurSystem \vs prom. gap } \\
    \end{array}
    \]
    \vspace{-0.2in} 
    \caption{Viability of the approximations in
      \OurSystem. See discussion in
      \Sec{eval:ablation}. The graphs show the
      relative errors of (a) {\ct req-to-req} over
      {\ct req-to-res}; (ii) {\ct prominent-gap}
      over {\ct req-to-req}; and (iii) \OurSystem
      over {\ct prominent-gap}, at three pipeline
      depths.}
    \label{fig:ablation}
    \end{minipage}
    \hfill %
    \begin{minipage}[b]{0.23\textwidth}
    \centering
    \[
    \begin{array}{c}
    \includegraphics[width=\linewidth]{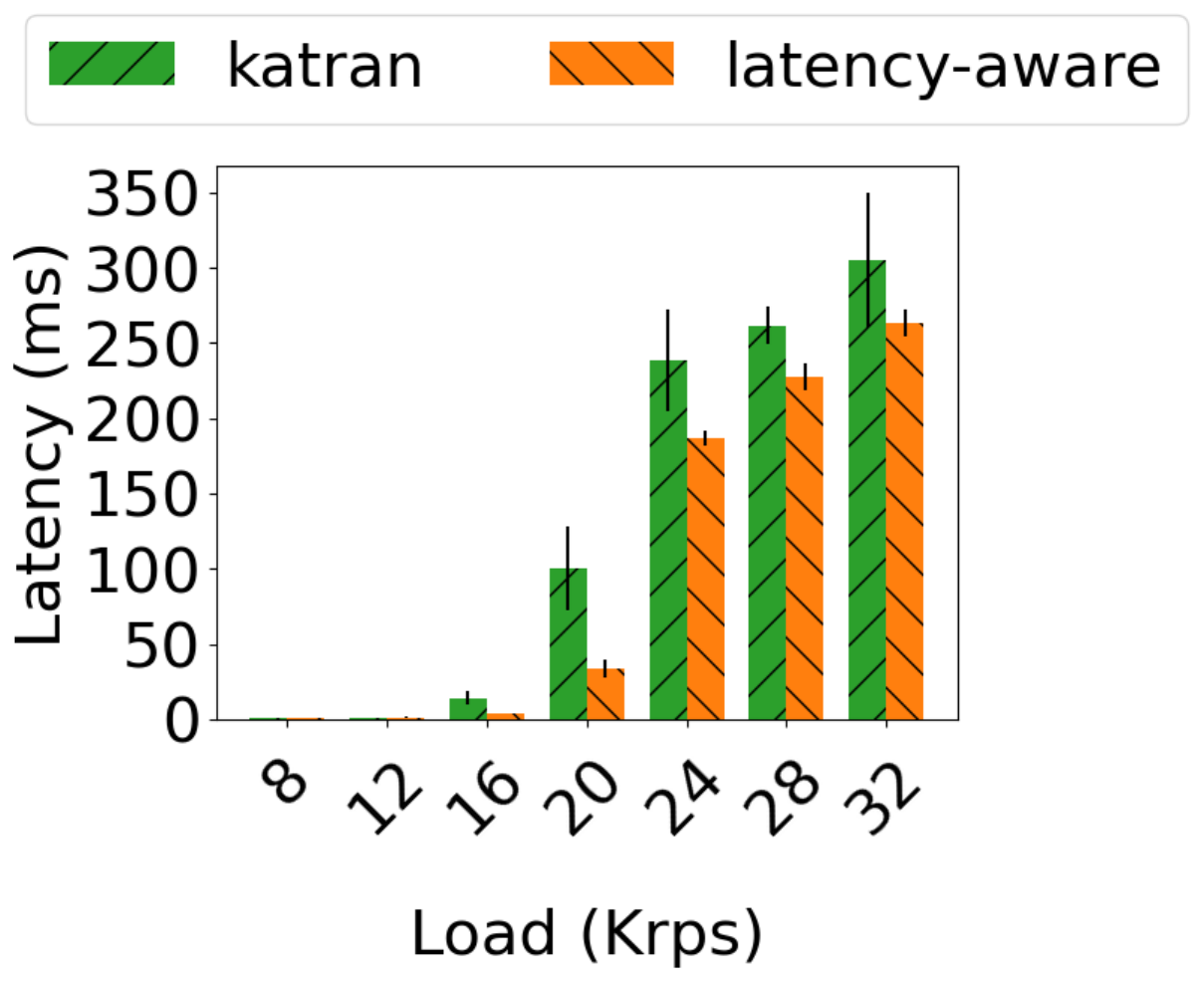}\\
    \textrm{Feedback control} \\
    \end{array}
    \]
    \vspace{-0.2in}
    \caption{Comparing 99th percentile tail latency of Katran and latency-aware one
    (\Sec{eval:loadbalancer}).}
    \label{fig:loadbalancer}
    \end{minipage} 
\end{figure*}

\subsection{Ablation to Evaluate Key Ideas}
\label{sec:eval:ablation}
We assess the core ideas in \OurSystem through an ablation study, 
focusing on three main questions: (i) Does the
time gap between causally-related packets reflect
the application-level response latency
(\Sec{causal-idea})? (ii) How likely is a packet
following a prominent time gap (\ie, a time gap
``significantly'' larger than other inter-packet
time gaps) to be a \ctp
(\Sec{packet-gap-assumption})?  (iii) Are
prominent time gaps captured accurately by
maintaining distributions in \OurSystem
(\Sec{estimating-packet-gap})?

We run our RPC benchmarking client and server
(\Sec{eval:accuracy}) within a single cluster
(\Fig{experiment-topology}(a)). To simplify the
experiment, we leave the server performance
unconstrained (different from
\Sec{eval:accuracy}), and collect measurements
over one long-lived client-server connection. We
measure: (1) the ground truth response latency
({\ct req-to-res}), and also three successively
coarse approximations of it: (2) causal pair delay
({\ct req-to-req}, \Sec{causal-idea}); (3) an
approach that uses a threshold of 0.6 $\times$ the
true response latency, to separate packets into
batches of causally-related packets
(\Sec{packet-gap-assumption}), labeled {\ct
  prominent-gap}; and (4) \OurSystem
(\Sec{estimating-packet-gap}).
%
%
In our experiments, the number of concurrent
requests in flight is a key factor that impacted
measurement accuracy, hence we evaluate three
pipeline depths (4, 8, 16).

\Fig{ablation}(a), (b), and (c) show the CDF of
the relative error of each successively
approximate method listed above with respect to
the method before it.  We think that the overall
quality of the approximations remains high across
settings. The errors in the tail worsen as
pipeline depth increases. 

The resilience of \oursystem in relation to the offered load, packet loss, 
and packet reordering is detailed in \App{robustness}. Furthermore, the findings 
regarding the efficacy of the heuristics are documented in \App{heuristics}.

\nop{
We now present the results for validity of the
claim that request to triggered request is a good
approximation for request to response. In this
experiment, we use a threshold gap that is
representative of idle period.  A time interval
between two batch of packets separated by this
threshold gap is emitted as a latency sample.  We
compare this latency sample with the
request-to-request and request-to-response
collected at the client. In \Fig{ablation} P4, P8,
P16 is the comparison of request-to-request
observed at the client against request-to-response
observed at the client for pipeline depth of 4, 8
and 16. The remaining 6 curves compares the common
samples collected by threshold based approach and
request-to-request to its request-to-response
counterpart. P4, P8, P16 includes samples for all
the requests whereas remaining curves represents a
subset of P4,P8 and P16 as multiple requests can
be clubbed into a single packet. The threshold
based method deviates from request to response
with a large margin with pipeline depth of
16. This is due to staggered response as mentioned
in \Sec{eval:accuracy}
}

\vspace{-0.1in}

\subsection{Feedback Control}

\label{sec:eval:loadbalancer}



Can measurements from \OurSystem enable robust
real-time feedback control? To answer this
question, we evaluate the use of measurements from
\OurSystem within a layer-4 load balancer,
Katran~\cite{katran-description,
  katran-facebook-talk}. Katran implements direct
server return~\cite{ananta-sigcomm13} to bypass
the load balancer for server responses. We use the
controller discussed in
\Sec{using-response-latency-for-server-load-balancing}.

This experiment uses the star topology shown in
\Fig{experiment-topology}(c), where two clients
connect to four servers via a measurement vantage
point that uses the \OurSystem measurements to
balance server load.
The seven server machines include an Intel
E5-2640v4 10-core CPU, 64GB memory, and dual-port
25Gb NIC. A Dell S4048 switch interconnects the
machines. We configured the switch and all the
servers to forward clients-to-servers packets via
the vantage point, and server-to-client packets
directly via the switch, mimicking a direct server
return
configuration~\cite{katran-example-setup}. The
\oursystem algorithm runs on the switch-facing
ingress interface of the vantage point.
We simplified the web workload from
\Sec{eval:accuracy} so that each client connection
maintains only one request in flight. Server
performance is varied over time using a trace from
a production-grade compute cluster, as discussed
in \Sec{eval:accuracy}.
Similar to \Sec{eval:overheads}, \OurSystem
implements an eBPF tail call into the Katran
code~\cite{ebpf-tail-call}.

\Fig{loadbalancer} 
compares the 99th percentile
tail response latency of the latency-aware load
balancer with that of the vanilla version at varying
offered loads (Krequests/sec).  Latency awareness
produces, on average, a 37\% reduction in tail
latency across loads. Moreover, latency awareness
also results in more predictable tail latencies
(smaller error bars). 

%



\section{Discussion}
\label{sec:discussion}

This paper presented \oursystem, an algorithm that
passively and continuously measures response
latencies, remaining able to operate over
encrypted application and transport headers, and
under routing asymmetry. \OurSystem leverages the
idea of \cps, two requests whose latter is
triggered by the response to the former. In our
experiments, \OurSystem shows high accuracy and
practical software overheads. \OurSystem can
enable robust real-time feedback control.

\nop{
\OurSystem has two significant limitations, both
stemming from the estimator's assumptions. First,
applications with significant client think times
suffer from sizable overestimations of response
latency (\eg DASH video streams). Second, the
estimator assumes that successive batches of
packets are causally related. However, large
response objects (\Sec{eval:accuracy}), and the
presence of arbitrary reordering in server
responses (relative to requests) both violate this
assumption. The latter challenges generalizing our
results to HTTP/2 and QUIC.

Despite these limitations, \OurSystem is robust
for HTTP/1.1-like web and fast-growing API
traffic~\cite{cloudflare-api-traffic-explosion,
  cloudflare-http-api-usage}, in the presence of
multiple pipelined concurrent requests, server
fluctuations, and network variability. These
results may have broader implications to the
design of reactive feedback control systems
relying on passive continuous measurement.
}





\forcameraready{\section*{Acknowledgments}

Srikanth Kandula
Laurent Vanbever

}

\ifacm {
  \bibliographystyle{ACM-Reference-Format}
}
\ifusenix {
  \bibliographystyle{plain}
}
\bibliography{reference.bib}

\appendix
\section{An Estimator based on Prominent Packet
  Gaps}
\label{app:packet-gap-algorithm}

See \Alg{fixed-timeout-detection}.

\packetgapalgorithm

\section{Maintaining Efficient Histograms}
\label{app:histogram-algo}

See \Alg{efficient-mode-update}.

\efficientmodeupdatealgorithm

\section{Detailed Design of the Latency-Aware Load
  Balancer}
\label{app:detailed-lb}

Below, we present the details of the design of our
latency-aware layer-4 load balancer, which builds
on a Maglev-like load
balancer~\cite{maglev-nsdi16} which permits the
setting of weights to servers. The load balancer
should measure response latency for a number of
connections mapped to each server using the
algorithms in \Sec{design}, to produce a
representative average latency for each server. We
then use these average server latencies to adapt
the weights assigned to the servers, using three
key ideas.

First, we take away weights from servers which have average latencies
larger than a high watermark, and place them on servers which have
average latencies smaller than a low watermark. Our high
(respectively, low) latency watermark is defined as $\alpha_{high}$
(respectively, $\alpha_{low}$) times the latency of the server with
the smallest average latency. We use $\alpha_{high} = 1.5$ and
$\alpha_{low}= 1.2$. Moreover, the weight shifted from a high-latency
server is proportional to its latency.

Second, we restrict the low-latency servers eligible for placing
additional weight by limiting ourselves to servers that have
sufficient {\em freshness} in their latency measurement. As prior work
has observed~\cite{prequal-nsdi24}, measured latencies are a good
metric for the past performance of backend servers, but not their
future. Instead, a different metric such as the number of requests in
flight to a server is a leading indicator for the future performance
of that server. Consequently, some prior works consider a combination
of requests in flight and latency to balance
load~\cite{prequal-nsdi24, c3-nsdi15}.  A DSR load
balancer cannot directly measure the number of requests
in flight.

Instead, we measure the recency of the latency measurement of servers,
by defining the freshness of a server latency measurement to be the
ratio of the total number of requests received in the last measurement
interval to the number of concurrent active connections at the end of
that interval.  This metric captures the intuition that a server on
the verge of slowing down will have processed fewer requests per
active connection. Conversely, a fast server must have processed more
requests in the last measurement interval even if several of those
connections have arrived and completed. We only consider a low-latency
server eligible to take on additional weight if its freshness is at
least as high as any high-latency server. Each such server receives an
equal share of the total weight that is shifted away from the
high-latency servers, subject to a cap on the per-server increment in
weight, to avoid the thundering herd
problem~\cite{memcache-facebook-nsdi13}.

The third key idea is to regress to the mean: if latencies have not
changed in the recent $k$ measurement intervals, the weights are
slowly equalized across servers (we use $k=3$). Server and network
performance can be highly variable. It is faster to improve
performance from an operating point where no one server is assigned a
disproportionately large weight.

\section{Heuristics in \OurSystem}
\label{app:heuristics}

To further improve the accuracy of
\Alg{efficient-mode-update}, we use two heuristics
that eliminate \IPG modes that we deem
noisy. First, we coalesce \IPGs corresponding to
pure transport-layer acknowledgments into one
\IPG, by ignoring them from the stream of \IPGs
observed for a connection.  The intuition is that
pure ACKs do not represent the completion of an
application-layer response, but rather signal
partial completion. Mechanically, when transport
headers are unencrypted, it is easy to identify
pure ACK packets by inspecting the transport-layer
headers, for example, the TCP ACK flag and packet
size. When the transport layer is encrypted, this
heuristic only applies if there is information
that helps classify a packet as a pure ACK, \eg
payload sizes.
Our second heuristic is to explicitly mark \IPGs
following a non-MTU packet as candidate modes to
represent the \IBG used in the proportional mode sum
computation. The intuition is that clients typically
send full MTU packets whenever packets can be
transmitted independently and back to back.




\begin{figure}[ht!]
    \[
    \begin{array}{cc}
    \includegraphics[width=0.23\textwidth]{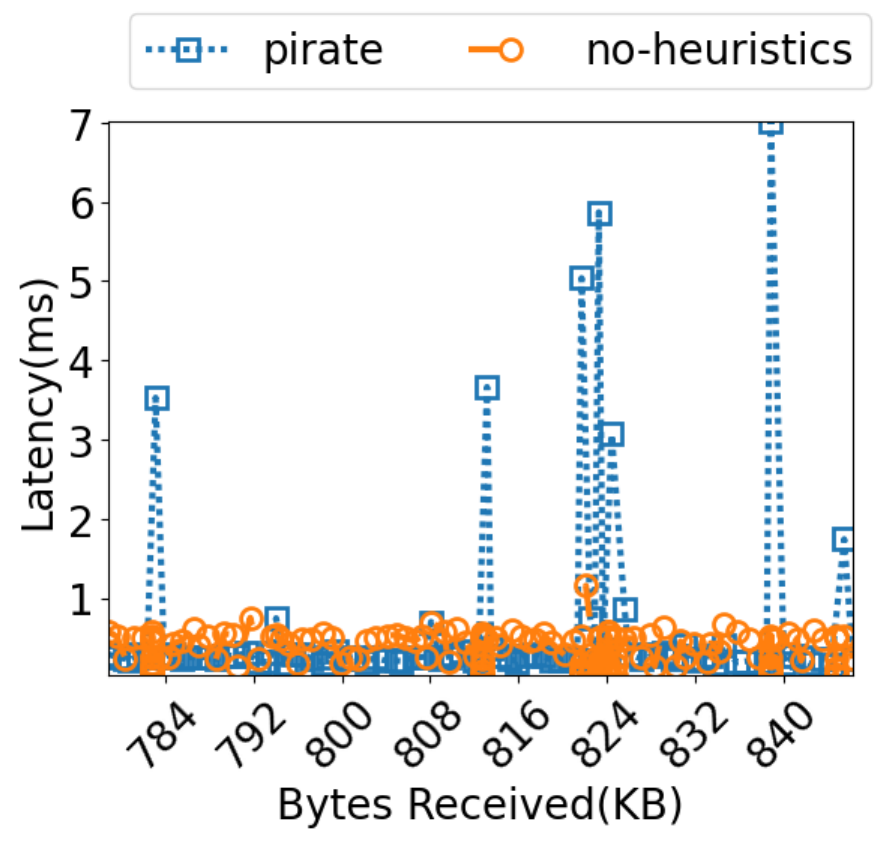} &
    \includegraphics[width=0.23\textwidth]{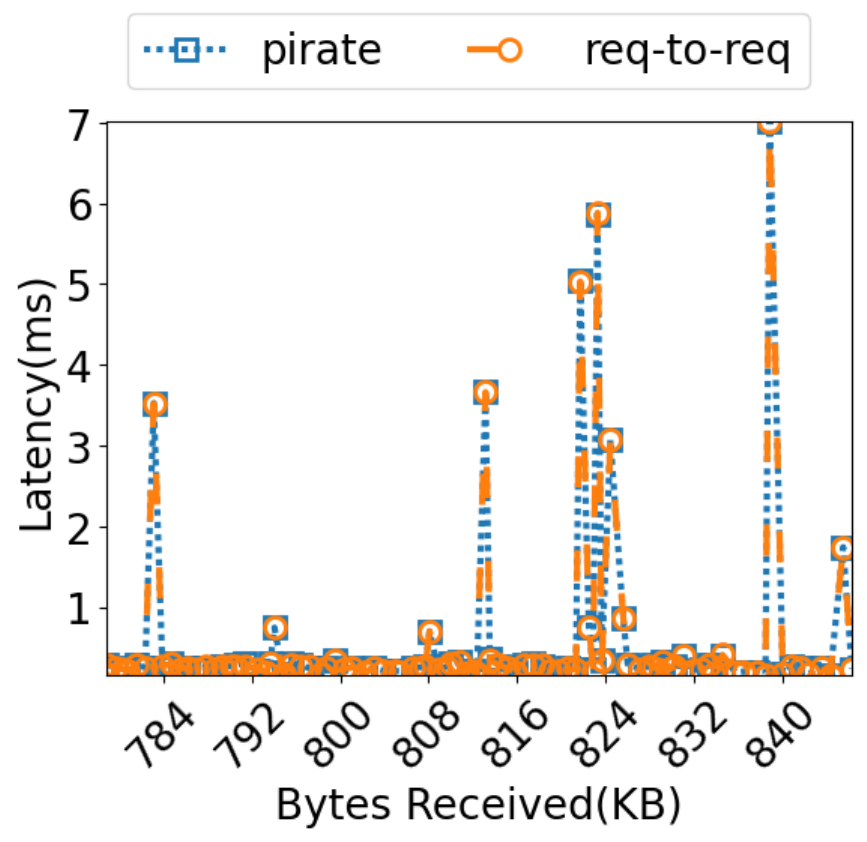} \\
    \end{array}
    \]
    \caption{
      (a) \oursystem \vs\ \oursystem without both heuristics.
      (b) \oursystem \vs the request-to-triggered request delay.}
    \label{fig:heuristics}
\end{figure}

Using our setup described in
\Sec{detailed-experimental-setup}, we evaluate the
benefits of the ACK-coalescing and MTU size heuristics
(\Sec{efficient-ipg-distributions}) in improving the
accuracy of \OurSystem's estimation.
\Fig{heuristics}(a) compares the estimated response
latencies when turning off both heuristics, against
\oursystem. In (b), the time series of
request-to-triggered-request delays is shown against
\oursystem for reference.
The heuristics provide a noticeable improvement to
accuracy, especially in allowing \oursystem to track
fast fluctuations in latency.

\nop{
When ACK coalescing is not
enabled the interpacket gaps due to ACKs create
additional modes that mix with the modes that
correspond to gaps between requests and triggered
requests.
%
%
\ngs{Text below needs to be clarified and cleaned up.}
When MTU heuristic is disabled, and the measurement
interval has multiple response latencies, the
computation of the proportional mode sum
(\Sec{empirical-probability-distribution}) returns a
higher latency value. By default, only the
highest-valued mode is assumed to be a prominent gap
\ngs{pointer to where we defined this earlier}, and all
other modes are assumed to be packet gaps that occur
prior to that gap. However, these multiple gaps do not
co-occur during the same response latency period, and
merely computing a proportional mode sum can lead to a
latency estimate that is significantly and erroneously
higher than the true average latency over the
measurement epoch. 
\Fig{heuristics}(c) compares the
estimated latency of pirate with the variant of
\oursystem with both heuristics disabled.
}

\section{Robustness of the Estimator}
\label{app:robustness}

\begin{figure}[ht]
    \centering
    \[
    \begin{array}{cc}
    \includegraphics[width=0.22\textwidth]{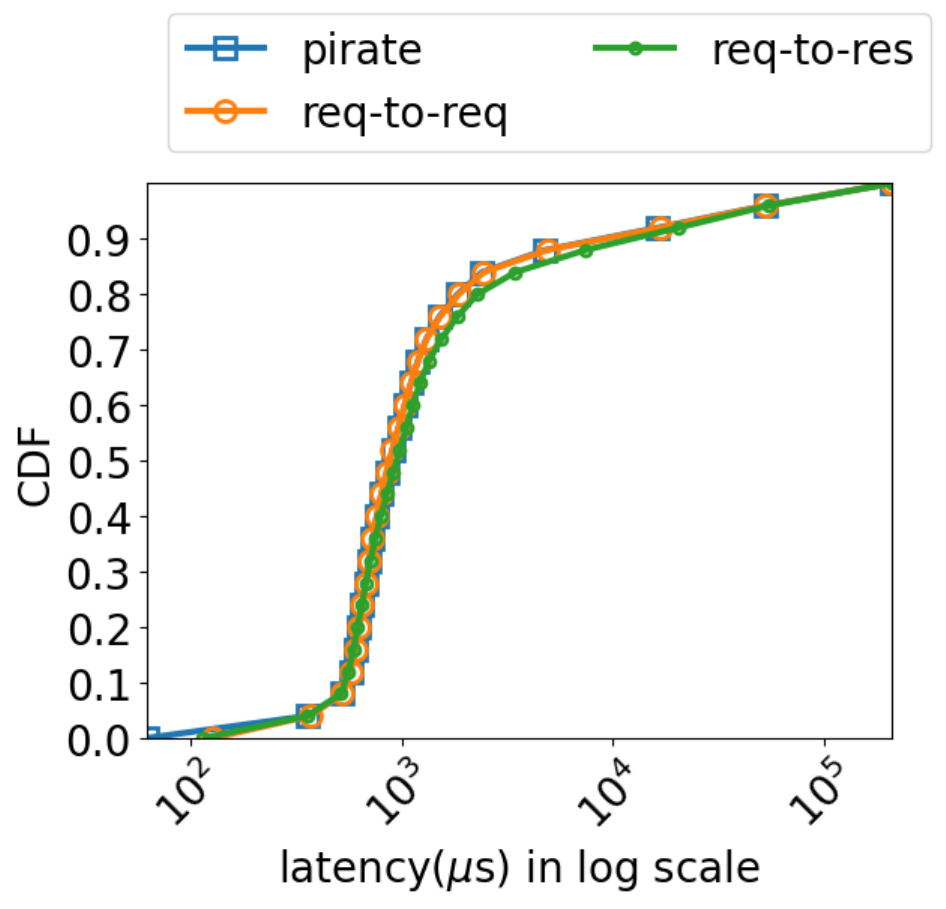} &
    \includegraphics[width=0.23\textwidth]{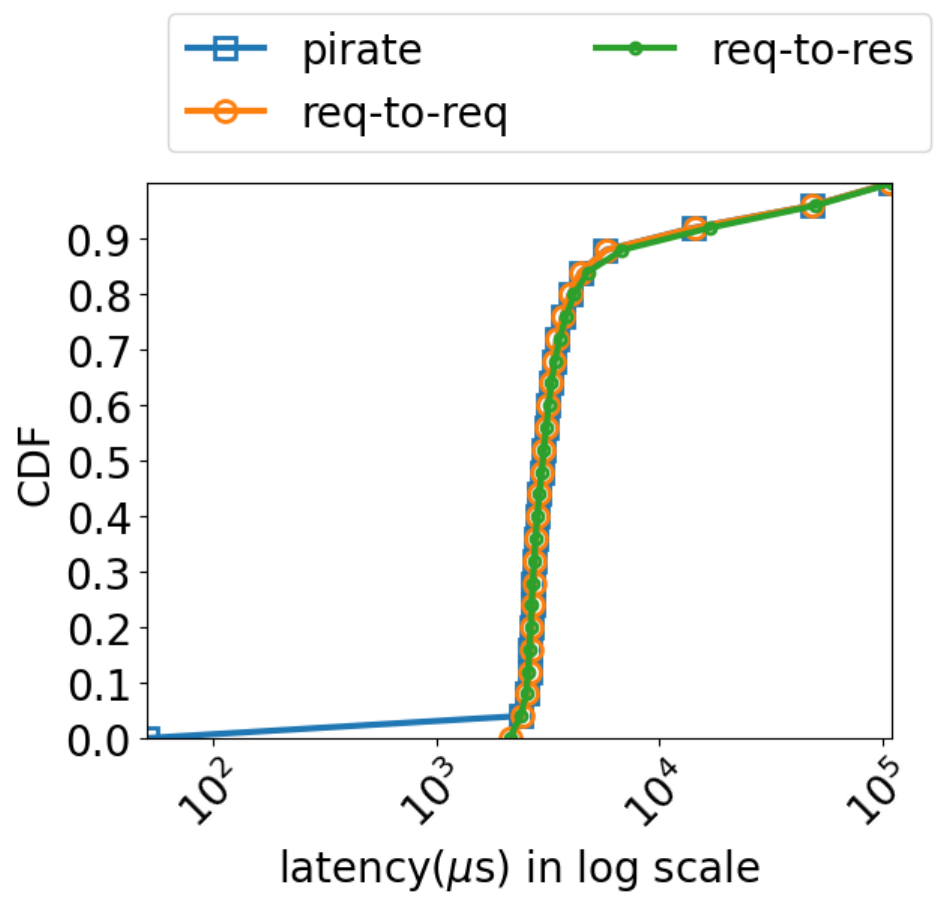}\\
    \end{array}
    \]
    \vspace{-0.2in}
    \caption{
      (a) CDF of latencies for a loss rate of 1\%
      (b) CDF of latencies for a packet reorder rate of 25\%}
    \label{fig:loss}
\end{figure}

\Para{Packet Loss and Reordering.} We evaluate the
robustness of \oursystem to packet loss and reordering,
which may affect packet inter-arrival times at the
vantage point due to the transport layer's
adaptation. We induce loss and reordering of varying
rates over packets from the client towards the vantage
point. (Loss on the server-to-client path triggers
retransmission from the server to the client, which is
not observed by the vantage point, hence we do not
evaluate this.)
%
%
\Fig{loss}(a) shows the CDF of latencies measured by
\OurSystem under a loss rate of 1\%, while
\Fig{loss}(b) shows the CDF of latencies under a high
packet reorder rate of 25\%. 
\OurSystem produces robust estimates of response
latency under both conditions. 
Estimation accuracies were similar under other loss and
reorder rates that we evaluated.
%

\begin{figure}[ht]
    \centering
    \[
    \begin{array}{cc}
    \includegraphics[width=0.23\textwidth]{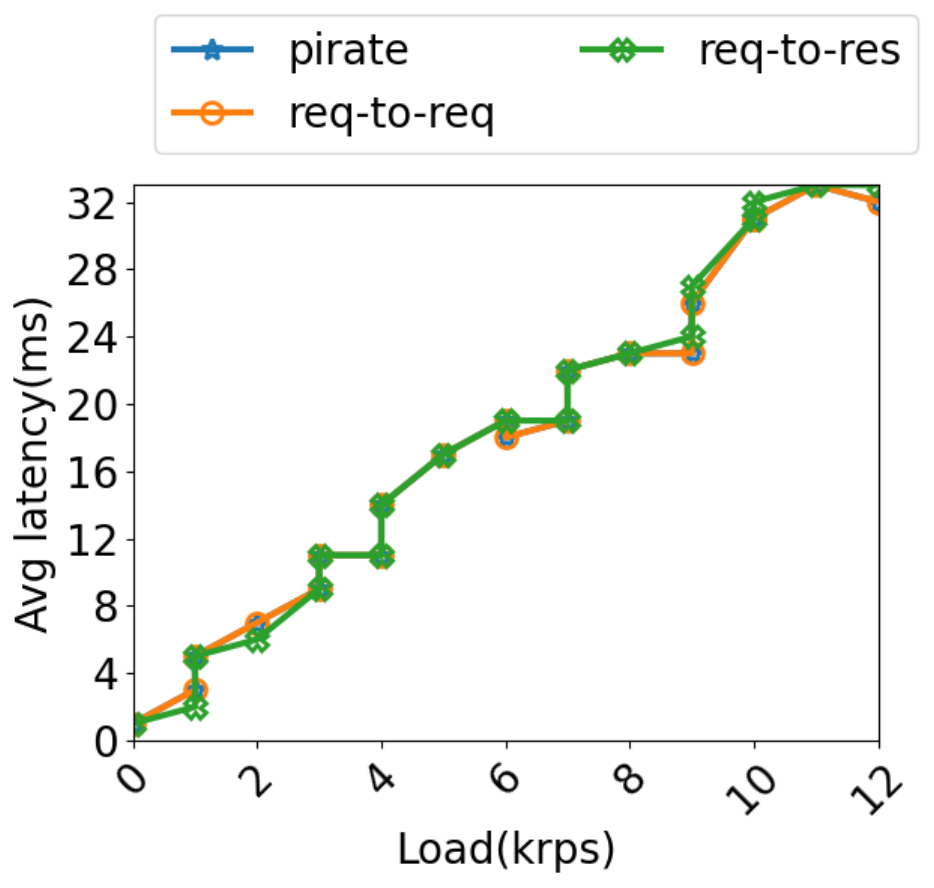} &
    \includegraphics[width=0.22\textwidth]{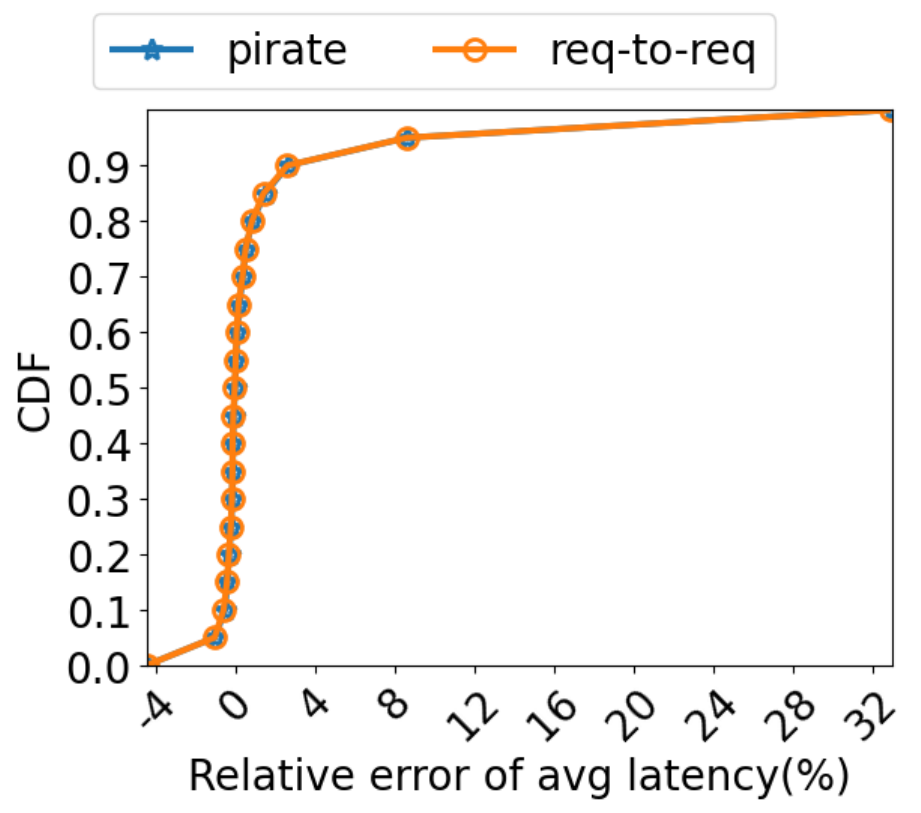} \\
    \end{array}
    \]
    \vspace{-0.2in}
    \caption{ (a) Average response latency across
      all connections vs. offered load.  (b)
      Relative error at 12K requests/sec.}
    \label{fig:request-comparison}
\end{figure}

\Para{Robustness to Offered Load.} We evaluated the
accuracy of latency estimation as the load offered to
the server varies. 
In this experiment, we use the simplified web workload
from \Sec{eval:comparison-tcp-rtt} to push the server
load to higher request/second values.
\Fig{request-comparison}(a) shows the average latency
across all connections at varying loads
(requests/second).
At a load of 12K requests/second, in
\Fig{request-comparison}(b), we show the CDF of
relative error (averaged per connection) across the
techniques.
In both cases, \OurSystem agrees strongly with the
response latency measured at the client.

\end{document}